\documentclass[dvips, a4paper, 12pt]{article}

\usepackage{indentfirst}

\usepackage[english]{babel}
\usepackage[latin1]{inputenc}

\usepackage{color}
\definecolor{oneblue}{rgb}{0,0.0,0.75}
\usepackage[colorlinks,
            urlcolor=oneblue,
            linkcolor=oneblue,
            citecolor=oneblue,
            bookmarksopen=false,
            pagebackref]{hyperref}

\usepackage[dvips,final]{graphicx}

\usepackage[sort]{natbib}

\usepackage{algorithm}
\usepackage{algorithmic}

\usepackage{xspace} 
\usepackage{amsmath,amssymb,amsthm}

\usepackage{fancyhdr}
\newtheorem{lemma}{Lemma}
\newcommand{\pd}[2]{\frac{\partial#1}{\partial#2}}
\newcommand{\abs}[1]{\left|#1\right|}
\newcommand{\set}[1]{\left\{ #1 \right\}}
\newcommand{\eps}{\varepsilon}
\newcommand{\pphi}{\varphi}
\newcommand{\m}{\mu^2}
\newcommand{\mm}{\mu^4}
\renewcommand{\Re}{\mathop{\mathrm{Re}}}
\renewcommand{\Im}{\mathop{\mathrm{Im}}}

\newcommand{\rot}{\mathop{\mathrm{rot}}}

\newcommand{\sech}{{\rm sech}\,}
\newcommand{\fft}{{\rm fft}\,}
\newcommand{\ifft}{{\rm ifft}\,}

\def\u{\mathbf{u}}
\def\k{\mathbf{k}}
\def\U{\mathbf{U}}
\def\L{\mathbf{L}}
\def\F{\mathbf{F}}
\def\D{\mathbf{D}}
\def\S{\mathbf{S}}
\def\P{\mathbf{P}}
\def\O{\mathbf{0}}
\def\R{\mathbb{R}}
\def\ua{\mathbf{u_\alpha}}
\def\div{\nabla\cdot}
\def\za{z_\alpha}
\title{Dissipative Boussinesq equations}
\author{\href{http://www.cmla.ens-cachan.fr/\~dutykh}{Denys Dutykh%
\footnote{CMLA, ENS Cachan, CNRS, PRES UniverSud, 61 Av. President Wilson, F-94230 Cachan, FRANCE}} \and
\href{http://www.cmla.ens-cachan.fr/\~dias}{Frédéric Dias\footnotemark[1]}}
\date{}

\begin{document}
\maketitle

\begin{abstract}
The classical theory of water waves is based on the theory of inviscid flows. However it is important to include viscous effects in some applications. Two models are proposed to add dissipative effects in the context of the Boussinesq equations, which include the effects of weak dispersion and nonlinearity in a shallow water framework. The dissipative Boussinesq equations are then integrated numerically. 
\end{abstract}

\tableofcontents

\section{Introduction}

Boussinesq equations are widely used in coastal and ocean engineering. One example among others is tsunami wave modelling. These equations can also be used to model tidal oscillations. Of course, these types of wave motion are perfectly described by the Navier-Stokes equations, but 
currently it is impossible to solve fully three-dimensional (3D) models in any significant domain.
Thus, approximate models such as the Boussinesq equations must be used.

The years 1871 and 1872 were particularly important in the development of the Boussinesq 
equations. It is in 1871 that Valentin Joseph Boussinesq received the Poncelet prize from the Academy of Sciences for his work. In the Volumes 72 and 73 of the ``Comptes Rendus Hebdomadaires des Séances de l'Académie des Sciences'', which cover respectively the six-month periods January--June 1871 and July--December 1871, there are several contributions of Boussinesq.
On June 19, 1871, Boussinesq presents the now famous note on the solitary wave entitled ``Théorie de l'intumescence liquide appelée onde solitaire ou de translation, se propageant dans un canal rectangulaire'' ({\bf 72}, pp. 755--759), which will be extended later in the note 
entitled ``Théorie générale des mouvements qui sont propagés dans un canal rectangulaire horizontal'' ({\bf 73}, pp. 256--260). Saint-Venant presents a couple of notes of Boussinesq entitled ``Sur le mouvement permanent varié de l'eau dans les tuyaux de conduite et dans les canaux découverts'' ({\bf 73}, pp. 34--38 and pp. 101--105). Saint-Venant himself
publishes a couple of notes entitled ``Théorie du mouvement non permanent des eaux, avec application aux crues des rivières et à l'introduction des marées dans leur lit'' ({\bf 73}, pp. 147--154 and pp. 237--240). All these notes deal with shallow-water
theory. On November 13, 1871, Boussinesq submits a paper entitled ``Théorie des ondes et des remous qui se propagent le long d'un canal rectangulaire horizontal, en communiquant au liquide contenu dans ce canal des vitesses sensiblement 
pareilles de la surface au fond'', which will be published in 1872 in the Journal de Mathématiques Pures et Appliquées
({\bf 17}, pp. 55--108). 

\cite{Boussinesq1871, Boussinesq1872} included dispersive effects
for the first time in the Saint-Venant equations \citep{SV1871}.
One should mention that Boussinesq's derivation was restricted to $1+1$ dimensions ($x$ and $t$) and a horizontal bottom.
Boussinesq equations contain more physics than the Saint-Venant equations
but at the same time they are more complicated from the mathematical and numerical 
point of views. These equations possess a hyperbolic structure (the same as in the nonlinear 
shallow-water equations) combined with high-order derivatives 
to model wave dispersion. There have been a lot of further developments of these equations like 
in \cite{Peregrine1967, Nwogu1993, Wei1995, Madsen1998}. 

Let us outline the physical assumptions. The Boussinesq equations are intended to describe the
irrotational motion of an incompressible homogeneous inviscid fluid in the long wave 
limit. The goal of this type of modelling is to reduce 3D problems to two-dimensional (2D) ones. This is done
by assuming a polynomial (usually linear) vertical distribution of the flow field,
while taking into account non-hydrostatic effects. This is the principal physical
difference with the nonlinear shallow-water (NSW) equations.

There are a lot of forms of the Boussinesq equations. This diversity is due 
to different possibilities in the choice of the velocity variable. In most cases one chooses the
velocity at an arbitrary water level or the depth-averaged velocity vector. The resulting
model performance is highly sensitive to linear dispersion properties. The right choice
of the velocity variable can significantly improve the propagation of moderately long waves. A good
review is given by \cite{Kirby}.
There is another technique used by \cite{BCS}. Formally, one can transform higher-order
terms by invoking lower-order asymptotic relations. It provides an elegant way
to improve the properties of the linear dispersion relation and it gives a quite general mathematical
framework to study these systems.

The main purpose of this article is to include dissipative
effects in the Boussinesq equations. It is well-known that the effect of viscosity on free oscillatory waves on deep water was 
studied by \cite{Lamb1932}. What is less known is that Boussinesq himself studied this effect as well. Boussinesq wrote
three related papers in 1895 in the ``Comptes Rendus Hebdomadaires des Séances de l'Académie des Sciences'':
(i) ``Sur l'extinction graduelle de la houle de mer aux grandes distances de son lieu de production : 
formation des équations du problème'' ({\bf 120}, pp. 1381-1386), (ii)
``Lois de l'extinction de la houle en haute mer'' ({\bf 121}, pp. 15-20), (iii)
``Sur la manière dont se régularise au loin, en s'y réduisant à une houle simple, toute agitation confuse mais 
périodique des flots'' ({\bf 121}, pp. 85-88). It should be pointed out that the famous treatise on hydrodynamics 
by Lamb has six editions. The paragraphs on wave
damping are not present in the first edition (1879) while they are present in the third edition (1906). The authors did not
have access to the second edition (1895), so it is possible that Boussinesq and Lamb published similar results at the same time.
Indeed Lamb derived the decay rate of the linear wave amplitude in two different ways: in paragraph 348 of the sixth edition by a 
dissipation calculation (this is also what \cite{Boussinesq1895} did) and in paragraph 349 by a direct calculation based on the 
linearized Navier-Stokes equations. Let $\alpha$ denote the wave amplitude, $\nu$ the kinematic viscosity of the fluid and $k$ the 
wavenumber of the decaying wave. Boussinesq (see Eq. (12) in \cite{Boussinesq1895}) and Lamb both showed that
\begin{equation}\label{Lamb_decay}
\frac{\partial\alpha}{\partial t} = -2\nu k^2 \alpha. 
\end{equation}
Equation (\ref{Lamb_decay}) leads to the classical law for viscous decay of waves of amplitude $\alpha$, namely
$\alpha \sim \exp(-2\nu k^2 t)$ (see Eq. (13) in \cite{Boussinesq1895} after a few calculations).

In the present paper, we use two different models for dissipation and derive corresponding systems of
long-wave equations. There are several methods to derive the
Boussinesq equations but the resulting equations are not the same.
So one expects the solutions to be different. We will investigate 
numerically whether corresponding solutions remain close or not.

One may ask why dissipation is needed in Boussinesq equations.
First of all, real world liquids are viscous. This physical effect is ``translated''
in the language of partial differential equations by dissipative terms (e.g. the Laplacian in the Navier-Stokes equations).
So, it is natural to have analogous terms in the long wave limit. In other words, a
non-dissipative model means that there is no energy loss, which is not pertinent
from a physical point of view, since any flow is accompanied by energy dissipation.

Let us mention an earlier numerical and experimental study by \cite{Bona1981}. They 
pointed out the importance of dissipative effects for accurate long wave modelling.
In the ``Résumé'' section one can read
\begin{quote}
[...] it was found that the inclusion of a dissipative
term was much more important than the inclusion of the nonlinear term, although
the inclusion of the nonlinear term was undoubtedly beneficial in describing the
observations [...].
\end{quote}
The complexity 
of the mathematical equations due to the inclusion of this term is negligible 
compared to the benefit of a better physical description.

Let us consider the incompressible Navier-Stokes (N-S) equations for a Newtonian fluid:
\begin{align*}
  \nabla\cdot \u^* &= 0, \\
  \pd{\u^*}{t^*} + \u^*\cdot \nabla\u^* &= -\frac{\nabla p^*}{\rho} + \nu\Delta\u^* + \frac{\F^*}{\rho},
\end{align*}
where $\u^*(x,y,z,t) = (u^*,v^*,w^*)(x^*,y^*,z^*,t^*)$ is the fluid velocity vector, $p^*$ the pressure, 
$\F^*$ the body force vector, $\rho$ the constant fluid density and $\nu$ 
the kinematic viscosity.

Switching to dimensionless variables by introducing a characteristic velocity $U$, a 
characteristic length $L$ and a characteristic pressure $\rho U^2$, neglecting body forces\footnote{The presence or absence of body forces is not important for discussing viscous effects.} in this discussion, 
the N-S equations become
\begin{align*}
  \nabla\cdot \u &= 0, \\
  \pd{\u}{t} + \u\cdot \nabla\u &= -\nabla p + \frac1{Re}\Delta\u,
\end{align*}
where $Re$ is the well-known dimensionless parameter known as the Reynolds number and defined as
\begin{equation*}
  Re=\frac{F_{inertia}}{F_{viscous}} = \frac{UL}{\nu}.
\end{equation*}

From a physical point of view the Reynolds number is a measure of the relative importance of inertial forces
compared to viscous effects. For typical tsunami propagation applications the characteristic particle velocity $U$ is about 
$5$ cm/s and the characteristic wave amplitude, which we use here as characteristic length scale, is about $1$ m.
The kinematic viscosity $\nu$ depends on the temperature but its order of magnitude for water 
is 10$^{-6}$ m$^2$/s. Considering that as the tsunami approaches the coast both the particle velocity and the wave amplitude
increase, one can write that the corresponding Reynolds number is of the order of
10$^{5}$ or 10$^{6}$\label{page:re}. This simple estimate clearly
shows that the flow is turbulent (as many other flows in nature).

It is a common practice in fluid dynamics (addition of an ``eddy viscosity'' into the governing equations
for Large Eddy Simulations\footnote{Boussinesq himself introduced the concept of eddy viscosity in his famous 680 page
paper entitled ``Essai sur la théorie des eaux courantes'' \citep{Boussinesq1877}.}) to ignore the small-scale vortices when one is only 
interested in large-scale motion. It can significantly simplify computational and modelling
aspects. So the inclusion of dissipation can be viewed as the simplest way to take into account 
the turbulence.

There are several authors \citep{Tuck, LH1992, SVBM2002, Skandrani, ddz, Ruvinsky1991} who included dissipation due to viscosity 
in potential flow solutions and there are also authors \citep{Kennedy2000, Zelt1991, Heitner1970} who already
included in Boussinesq models ad-hoc dissipative terms into momentum conservation equations
in order to model wave breaking. Modelling this effect is not the primary goal of the present paper, 
since the flow is no longer irrotational after wave breaking. Strictly speaking the 
Boussinesq equations can no longer be valid at this stage. Nevertheless scientists and engineers continue
to use these equations even to model the run-up on the beach.
In our approach a suitable choice of the eddy viscosity, which is a function 
of both space and time, can model wave breaking at least as well as in the articles cited above.

\section{Derivation of the Boussinesq equations}\label{sec:section1}

\begin{figure}[htbp]
	\centering
		\includegraphics[width=0.95\textwidth]{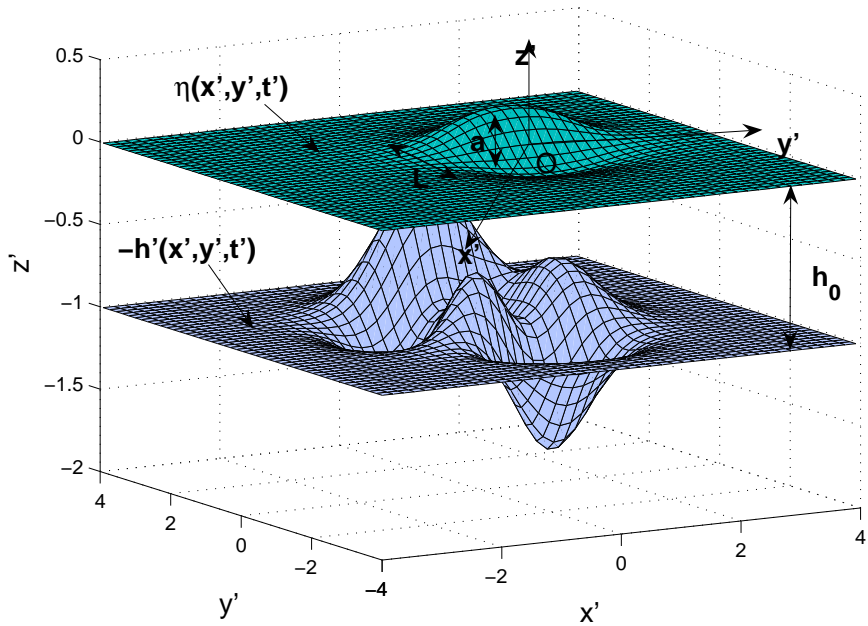}
	\caption{Sketch of the fluid domain}
	\label{fig:fluiddomain}
\end{figure}

In order to derive the Boussinesq equations, we begin with the full water-wave problem.
A Cartesian coordinate system $(x',y',z')$ is used, with the $x'-$ and $y'-$axis along the still water
level and the $z'-$ axis pointing vertically upwards.
Let $\Omega_t$ be the fluid domain in $\R^3$ which is occupied by an inviscid and 
incompressible fluid. The subscript $t$ underlines the fact that the domain varies with
time and is not known a priori. The domain $\Omega_t$ is bounded below by the seabed 
$z'=-h'(x',y',t')$ and above by the free surface $z'=\eta'(x',y',t')$. In this 
section we choose the domain $\Omega_t$ to be unbounded in the horizontal directions
in order to avoid the discussion on lateral boundary conditions. The reason is twofold. 
First of all, the choice of the boundary value problem (BVP)
(e.g. generating and/or absorbing boundary, wall, run-up on a beach) depends on the
application under consideration and secondly,
the question of the well-posedness of the BVP for the Boussinesq equations is essentially open.
Primes stand for dimensional variables. A typical sketch of the domain $\Omega_t$
is given in Figure \ref{fig:fluiddomain}. If the flow is assumed to be
irrotational one can introduce the velocity potential $\phi'$ defined by
$$
  \vec{u}' = \nabla' \phi', \qquad\nabla'\ := \left(\pd{}{x'},\pd{}{y'},\pd{}{z'}\right)^T,
$$
where $\vec{u}'$ denotes the velocity field. Then we write down the following 
system of equations for potential flow theory in the presence of a free surface:

\begin{equation*}
  \Delta' \phi' = 0, \qquad (x',y',z') \in \Omega_t = \R^2\times [-h', \eta'],
\end{equation*}
\begin{equation*}
  \phi'_{z'} = \eta'_{t'} + \nabla'\phi'\cdot\nabla'\eta', \qquad z' = \eta',
\end{equation*}
\begin{equation}\label{eq:bernoulli}
  \phi'_{t'} + \frac12\abs{\nabla'\phi'}^2 + g\eta' = 0, \qquad z' = \eta',
\end{equation}
\begin{equation*}
  \phi'_{z'} + h'_{t'} + \nabla'\phi'\cdot\nabla' h' = 0, \qquad z' = -h',
\end{equation*}
where $g$ denotes the acceleration due to gravity (surface tension effects are usually 
neglected for long-wave applications). It has been assumed implicitly that the free surface
is a graph and that the pressure is constant on the free surface (no forcing). Moreover
we assume that the total water depth remains positive, i.e. $\eta'+h'>0$ (there is no dry zone). 

As written, this system of equations does not contain any dissipation.
Thus, we complete the free-surface dynamic boundary condition (\ref{eq:bernoulli})
by adding a dissipative term to account for the viscous effects\footnote{\cite{ddz}, who considered deep-water waves,
pointed out that a viscous correction
should also be added to the kinematic boundary condition if one takes into account the vortical component of the
velocity. This correction was recently added in finite depth as well \citep{DD2007}. A boundary-layer correction
at the bottom was also included.}:
\begin{equation*}
  \phi'_{t'} + \frac12\abs{\nabla'\phi'}^2 + g\eta' + D'_{\phi'} = 0, \qquad z' = \eta'.
\end{equation*}

In this work we investigate two models for the dissipative term $D'_{\phi'}$. For simplicity,
one can choose a constant dissipation model (referred hereafter as 
Model I) which is often used (e.g. \citep{Jiang1996}):
\begin{equation}\label{eq:modelI}
  \mbox{Model I:} \qquad D'_{\phi'} := \delta_1 \phi'.
\end{equation}
There is a more physically realistic dissipation model which is obtained upon
balancing of normal stress on the free surface (e.g. \citep{Ruvinsky1991, Zhang1997, ddz}):
\begin{equation}\label{eq:modelII}
  \mbox{Model II:} \qquad D'_{\phi'} := \delta_2 \phi'_{z'z'}.
\end{equation}

The derivation of Boussinesq equations is more transparent when one works
with scaled variables. Let us introduce the following independent and dependent 
non-dimensional variables:
\begin{equation*}
  x = \frac{x'}{\ell},\qquad y = \frac{y'}{\ell},\qquad 
  z = \frac{z'}{h_0},\qquad t = \frac{\sqrt{gh_0}}{\ell}t',
\end{equation*}
\begin{equation*}
  h = \frac{h'}{h_0}, \qquad \eta = \frac{\eta'}{a_0}, \qquad
  \phi = \frac{\sqrt{gh_0}}{ga_0\ell}\phi',
\end{equation*}
where $h_0$, $\ell$ and $a_0$ denote a characteristic water depth, wavelength and
wave amplitude, respectively.

After this change of variables, the set of equations becomes
\begin{equation}\label{eq:conteq}
  \mu^2 (\phi_{xx} + \phi_{yy}) + \phi_{zz} = 0, \qquad (x,y,z)\in \Omega_t,
\end{equation}
\begin{equation}\label{eq:freesurfkin}
  \phi_z = \m\eta_t + \eps\m\nabla\phi\cdot\nabla\eta, \qquad z = \eps\eta,
\end{equation}
\begin{equation}\label{eq:dynbound}
  \m\phi_t + \frac12\eps\m\abs{\nabla\phi}^2 + \frac12\eps\phi_z^2 
    + \m\eta + \eps D_\phi = 0, \qquad z = \eps\eta
\end{equation}
\begin{equation}\label{eq:kinembound}
  \phi_z + \frac{\m}{\eps} h_t + \m\nabla\phi\cdot\nabla h = 0, \qquad z = -h,
\end{equation}
where $\eps$ and $\mu$ are the classical nonlinearity and frequency
dispersion parameters defined by
\begin{equation*}
  \eps := \frac{a_0}{h_0}, \qquad \mu := \frac{h_0}{\ell}.
\end{equation*}
In these equations and hereafter the symbol $\nabla$ denotes the horizontal gradient:
$$
  \nabla := \left(\pd{}{x},\pd{}{y}\right)^T.
$$
The dissipative term $D_\phi$ is given by the chosen model 
(\ref{eq:modelI}) or (\ref{eq:modelII}):
\begin{equation*}
  \mbox{Model I:} \;\; D_\phi = \frac{1}{R_1}\phi, \quad
  \mbox{Model II:} \;\; D_\phi = \frac{1}{R_2}\phi_{zz},
\end{equation*}
where the following dimensionless numbers have been introduced:
\begin{equation*}
  R_1 := \frac{1}{\delta_1}\left(\frac{ga_0\ell}{h_0^2\sqrt{gh_0}}\right), \quad
  R_2 := \frac{1}{\delta_2}\left(\frac{ga_0\ell}{\sqrt{gh_0}}\right).
\end{equation*}
From this dimensional analysis,
one can conclude that the dimension of the coefficient $\delta_1$ is $[s^{-1}]$ and that of $\delta_2$ is
$[m^2 s^{-1}]$. Thus, it is natural to call the first coefficient viscous frequency
(since it has the dimensions of a frequency) and the second one kinematic viscosity (by analogy with the N-S equations).

\begin{table}
\begin{center}
\begin{tabular}{lc}
  \hline
{\it parameter} & {\it value} \\
\hline \hline
  Acceleration due to gravity $g$, m/s$^2$ & 10 \\
  Amplitude $a_0$, m & 1 \\
  Wave length $\ell$, km & 100 \\
  Water depth $h_0$, km & 4 \\
  Kinematic viscosity $\delta$, m$^2$/s & $10^{-6}$ \\
\hline
\end{tabular}
\caption[]{Typical values of characteristic parameters in tsunami applications}
\label{tab:params}
\end{center}
\end{table}

It is interesting to estimate $R_2$, since
we know how to relate the value of $\delta_2$ to the kinematic viscosity of water.
Typical parameters which are used in tsunami wave modelling are given 
in Table \ref{tab:params}. For these parameters $R_2 = 5\times 10^9$ and
$\mu^2 = 1.6\times 10^{-3}$. The ratio between inertial forces and viscous forces is
$\frac12\eps\mu^2|\nabla\phi|^2/\eps|D_\phi|$. Its order of magnitude is $\mu^2 R_2$,
that is $8\times 10^6$. It clearly shows that the flow 
is turbulent and eddy-viscosity type approaches should be used. It means that,
at zeroth-order approximation, the main effect of turbulence is 
energy dissipation. Thus, one needs to increase the importance of viscous terms in
the governing equations in order to account for turbulent dissipation.

As an example, we refer one more time to the work by \cite{Bona1981}. They modeled
long wave propagation by using a modified dissipative Korteweg--de Vries equation:
\begin{equation}\label{eq:mKdV}
	  \eta_t + \eta_x + \frac32\eta\eta_x - \mu\eta_{xx} - \frac16\eta_{xxt} = 0.
\end{equation}
In numerical computations the authors took the coefficient $\mu = 0.014$. 
This value gave good agreement with laboratory data.

From now on, we will use the notation $\nu_i := 1/R_i$. This will allow us
to unify the physical origin of the numbers $R_i$ with the eddy-viscosity approach.
In other words, for the sake of convenience, we will ``forget'' about the origin of these coefficients, 
because their values can be given by other physical considerations.

\subsection{Asymptotic expansion}

Consider a formal asymptotic expansion of the velocity potential $\phi$ in powers
of the small parameter $\mu^2$:
\begin{equation}\label{eq:phiexp}
  \phi = \phi_0 + \m\phi_1 + \mm\phi_2 + \ldots\,.
\end{equation}

Then substitute this expansion into the continuity equation (\ref{eq:conteq})
and the boundary conditions. After substitution, the Laplace equation becomes
\begin{equation*}
  \m(\nabla^2\phi_0 + \m\nabla^2\phi_1 + \mm\nabla^2\phi_2 + \ldots) +
  \phi_{0zz} + \m\phi_{1zz} + \mm\phi_{2zz} + \ldots = 0.
\end{equation*}
Collecting the same order terms yields the following equations in the domain $\Omega_t$:
\begin{eqnarray}
  \mu^0: & \phi_{0zz} = 0, \label{eq:laplmu0} \\
  \m: & \phi_{1zz} + \nabla^2\phi_0 = 0, \label{eq:laplmu2} \\
  \mm: & \phi_{2zz} + \nabla^2\phi_1 = 0. \label{eq:laplmu4}
\end{eqnarray}
Performing the same computation for the bottom boundary condition yields the
following relations at $z=-h$:
\begin{eqnarray}
  \mu^0: & \phi_{0z} = 0, \label{eq:botmu0} \\
  \m: & \phi_{1z} + \frac1{\eps}h_t + \nabla\phi_0\cdot\nabla h = 0, \label{eq:botmu2} \\
  \mm: & \phi_{2z} + \nabla\phi_1\cdot\nabla h = 0. \label{eq:botmu4}
\end{eqnarray}
From equation (\ref{eq:laplmu0}) and the boundary condition (\ref{eq:botmu0}) one 
immediately concludes that
\begin{equation*}
  \phi_0 = \phi_0 (x,y,t).
\end{equation*}
Let us define the horizontal velocity vector
\begin{equation*}
  \u (x,y,t) := \nabla \phi_0, \quad \u = (u,v)^T.
\end{equation*}

The expansion of Laplace equation in powers of $\m$ gives recurrence relations between $\phi_0$, $\phi_1$,
$\phi_2$, etc. Using (\ref{eq:laplmu2}) one can express $\phi_1$ in terms of the derivatives of $\phi_0$:
\begin{equation*}
 \phi_{1zz} = -\nabla\cdot\u.
\end{equation*}
Integrating once with respect to $z$ yields
\begin{equation*}
  \phi_{1z} = -z\nabla\cdot\u + C_1 (x,y,t).
\end{equation*}
The unknown function $C_1(x,y,t)$ can be found by using condition (\ref{eq:botmu2}):
\begin{equation*}
  \phi_{1z} = -(z+h)\nabla\cdot\u - \frac{1}{\eps}h_t - \u\cdot\nabla h,
\end{equation*}
and integrating one more time with respect to $z$ gives the expression for $\phi_1$:
\begin{equation}\label{eq:phi1}
  \phi_1 = -\frac12(z+h)^2\nabla\cdot\u - z\left(\frac{1}{\eps}h_t + \u\cdot\nabla h\right).
\end{equation}

Now we will determine $\phi_2$. For this purpose we use equation (\ref{eq:laplmu4}):
\begin{multline}
  \phi_{2zz} = \frac12(z+h)^2 \nabla^2(\nabla\cdot\u) +
  \bigl((h+z)\nabla^2 h + \abs{\nabla h}^2\bigr)\nabla\cdot\u + \\ +
  2(h+z)\nabla h\cdot \nabla(\nabla\cdot\u) + 
  z\left(\frac{1}{\eps}\nabla^2 h_t + \nabla^2(\u\cdot\nabla h) \right).
\end{multline}

Integrating twice with respect to $z$ and using the bottom boundary condition (\ref{eq:botmu4}) 
yields the following expression for $\phi_2$:
\begin{multline}\label{eq:phi2}
  \phi_2 = \frac{1}{24}(h+z)^4 \nabla^2(\nabla\cdot\u) +
  \Bigl(\frac16(z+h)^3\nabla^2h + \frac12z^2\abs{\nabla h}^2\Bigr)\nabla\cdot\u \\ +
  \frac13(z+h)^3 \nabla h\cdot\nabla(\nabla\cdot\u) + 
  \frac{z^3}{6}\bigl(\frac{1}{\eps}\nabla^2 h_t + \nabla^2(\u\cdot\nabla h)\bigr) \\
  -zh\Bigl(\frac{h}{2}\nabla^2\bigl(\frac{1}{\eps}h_t + \u\cdot\nabla h\bigr) +
  \nabla h\cdot\nabla\bigl(\frac{1}{\eps}h_t + \u\cdot\nabla h\bigr)
  -\abs{\nabla h}^2\nabla\cdot\u\Bigr).
\end{multline}

\textbf{Remark:} In these equations one finds the term $(1/\eps)h_t$ due to the moving bathymetry. We would like to emphasize
that this term is $O(1)$, since in problems of wave generation by a moving bottom 
the bathymetry $h(x,y,t)$ has the following special form in dimensionless variables:
\begin{equation}\label{eq:defzeta}
  h(x,y,t) := h_0(x,y) - \eps\zeta(x,y,t),
\end{equation}
where $h_0(x,y)$ is the static seabed and $\zeta(x,y,t)$ is the dynamic
component due to a seismic event or a landslide (see for example \cite{Dutykh2006} for a practical algorithm
constructing $\zeta(x,y,t)$ in the absence of a dynamic source model). The amplitude
of the bottom motion has to be of the same order of magnitude as the resulting waves,
since we assume the fluid to be inviscid and incompressible.
Thus $(1/\eps)h_t = -\zeta_t = O(1)$.

In the present study we restrict our attention to dispersion terms up to order
$O(\m)$. We will also assume that the Ursell-Stokes number \citep{Ursell1953} is $O(1)$:
\begin{equation*}
S:=\frac{\eps}{\m} = O(1).
\end{equation*}
This assumption implies that terms of order $O(\eps^2)$ and
$O(\eps\m)$ must be neglected, since \label{page:Stokes}
\begin{equation*}
  \eps^2 = S^2\mm = O(\mm), \qquad
  \eps\m = S\mm = O(\mm).
\end{equation*}
Of course, it is possible to obtain high-order Boussinesq equations. We decided not to
take this research direction.
For high-order asymptotic expansions we refer to \cite{Wei1995, Madsen1998}.
Recently, \citep{Benoit2006} performed a comparative study between fully-nonlinear equations \citep{Wei1995} and 
Boussinesq equations with optimized dispersion relation \citep{Nwogu1993}.
No substantial difference was revealed. 

Now, we are ready to derive dissipative Boussinesq equations in their simplest form.
First of all, we substitute the asymptotic expansion (\ref{eq:phiexp}) into the
kinematic free-surface boundary condition (\ref{eq:freesurfkin}):
\begin{equation}\label{eq:almostetat}
  \phi_{0z} + \m\phi_{1z} + \mm\phi_{2z} = \m\eta_t + 
  \eps\m\nabla\phi_0\cdot\nabla\eta + O(\eps^2 + \eps\mm + \mu^6), 
  z = \eps\eta.
\end{equation}
The first term on the left hand side is equal to zero because of Eq. (\ref{eq:botmu0}).

Using expressions (\ref{eq:phi1}) and (\ref{eq:phi2}) one can evaluate $\phi_{1z}$ and
$\phi_{2z}$ on the free surface:
\begin{equation*}
  \left.\phi_{1z}\right|_{z=\eps\eta} = -(h+\eps\eta)\nabla\cdot\u - \frac{1}{\eps}h_t - \u\cdot\nabla h,
\end{equation*}
\begin{multline*}
 \left.\phi_{2z}\right|_{z=\eps\eta} = \frac{h^3}{6}\nabla^2(\div\u) + h^2\nabla h\cdot\nabla(\div\u) +
 h\left(\frac{h}{2}\nabla^2h + \abs{\nabla h}^2\right) \div\u \\
 -\frac{h^2}{2}\frac{1}{\eps}\nabla^2 h_t - h\frac1\eps\nabla h_t\cdot\nabla h + O(\eps).
\end{multline*} 
Substituting these expressions into (\ref{eq:almostetat}) and retaining only terms
of order $O(\eps + \m)$ yields the free-surface elevation equation:
\begin{multline*}
 \eta_t + \nabla\cdot\bigl((h+\eps\eta)\u\bigr) = -\left(1 + \frac{\m}{2}h^2\nabla^2 +
 \m h\nabla h\cdot\nabla\right)\frac{1}{\eps}h_t + \m\frac{h^3}{6}\nabla^2(\div\u) \\ +
 \m h\left(h\nabla h\cdot\nabla(\div\u) + \left(\frac{h}{2}\nabla^2h +
 \abs{\nabla h}^2\right)\div\u\right).
\end{multline*} 

The equation for the evolution of the velocity field is derived similarly from the dynamic
boundary condition (\ref{eq:dynbound}). This derivation will depend on the selected 
dissipation model. For both models one has to evaluate $\phi_1$, $\phi_{1t}$
and $\phi_{1zz}$ along the free surface $z = \eps\eta$ and then substitute the expressions into
the asymptotic form of (\ref{eq:dynbound}):
\begin{equation*}
  \m\phi_{0t} + \mm\phi_{1t} + \frac12\eps\m\abs{\nabla\phi_0}^2 + \m\eta + \eps\nu_2\m\phi_{1zz} = 
  O(\eps^2 + \eps\mm + \mu^6),
\end{equation*}
where, as an example, dissipative terms are given according to the second model. 
After performing all these operations one can write down the following equations:
\begin{eqnarray*}
  \mbox{Model I: } & \phi_{0t} + \frac{\eps}{2}\u^2 + \eta + \nu_1\frac{\eps}{\m}\phi_0
  - \frac{\nu_1\eps}2 h^2\nabla\cdot\u - \frac{\m}{2}h^2\nabla\cdot\u_t = 0, \\
  \mbox{Model II:} & \phi_{0t} + \frac{\eps}{2}\u^2 + \eta - \nu_2\eps\nabla\cdot\u 
  - \frac{\m}{2}h^2\nabla\cdot\u_t = 0.
\end{eqnarray*}
The last step consists in differentiating the above equations with respect to the
horizontal coordinates in order to obtain equations for the evolution of the velocity. We also perform some
minor transformations using the fact that the vector $\u$ is a gradient by definition,
so we have the obvious relation
\begin{equation*}
  \pd{u}{y} = \pd{v}{x}.
\end{equation*}

The resulting Boussinesq equations for the first and second dissipation models, 
respectively, are given below:
\begin{equation*}
 \eta_t + \nabla\cdot\bigl((h+\eps\eta)\u\bigr) = -\left(1 + \frac{\m}{2}h^2\nabla^2 +
 \m h\nabla h\cdot\nabla\right)\frac{1}{\eps}h_t + \m\frac{h^3}{6}\nabla^2(\div\u)
\end{equation*}
\begin{equation}\label{eq:etacommon} 
\hspace{3cm} + \m h\left(h\nabla h\cdot\nabla(\div\u) + \left(\frac{h}{2}\nabla^2h +
 \abs{\nabla h}^2\right)\div\u\right),
\end{equation} 
\begin{eqnarray}\label{eq:umodelI}
  \mbox{Model I:} & \u_t + \frac12\eps\nabla\u^2 + \nabla\eta + \nu_1 S\u = \frac12\eps\nu_1\nabla(h^2\div\u)
  + \frac12\m\nabla(h^2\div\u_t), \\
  \mbox{Model II:} & \u_t + \frac12\eps\nabla\u^2 + \nabla\eta  = \eps\nu_2\nabla^2\u + \frac12\m\nabla(h^2\div\u_t).\label{eq:umodelII}
\end{eqnarray}

\section{Analysis of the linear dispersion relations}

For simplicity, we will consider in this section only 2D
problems. The generalization to 3D problems is straightforward 
and does not change the analysis.

\subsection[Linear full potential flow equations with dissipation]{Linearization of the full potential flow equations with dissipation}
First we write down the linearization of the full potential flow equations in dimensional form, after
dropping the primes:
\begin{eqnarray}
  \Delta\phi = 0,& \qquad (x,z)\in \R\times[-h,0],\label{eq:lapllin} \\
  \phi_z = \eta_t,& \qquad z = 0, \label{eq:kinsurflin} \\
  \phi_t + g\eta + D_\phi = 0,& \qquad z = 0, \label{eq:dynsurflin} \\
  \phi_z = 0,& \qquad z = -h. \label{eq:botlin}
\end{eqnarray}

\textbf{Remark:} In this section the water layer is assumed to be of uniform depth,
so $h=$ const.

As above the term $D_\phi$ depends on the selected dissipation model and is equal to 
$\delta_1\phi$ or $\delta_2\phi_{zz}$. The next step consists in choosing a special
form of solutions:
\begin{equation}\label{eq:periodic}
  \phi(x,z,t) = \varphi_0 e^{i(kx-\omega t)}\pphi(z), \qquad
  \eta(x,t) = \eta_0 e^{i(kx-\omega t)},
\end{equation}
where $\varphi_0$ and $\eta_0$ are constants. Substituting this form
of solutions into equations (\ref{eq:lapllin}), (\ref{eq:kinsurflin})
and (\ref{eq:botlin}) yields the following boundary value problem for
an ordinary differential equation:
\begin{equation*}
  \pphi''(z) - k^2\varphi(z) = 0, \qquad z\in [-h, 0],
\end{equation*}
\begin{equation*}
  \pphi'(0) = \frac{\eta_0}{\varphi_0}(-i\omega), \qquad \pphi'(-h) = 0.
\end{equation*}
Straightforward computations give the solution to this problem:
\begin{equation*}
  \pphi(z) = -i\frac{\eta_0}{\varphi_0}
  \left(\frac{e^{k(2h+z)} + e^{-kz}}{e^{2kh}-1}\right)\frac{\omega}{k}.
\end{equation*}

The dispersion relation can be thought as a necessary condition for 
solutions of the form (\ref{eq:periodic}) to exist. The problem is that $\omega$
and $k$ cannot be arbitrary. We obtain the required relation $\omega = \omega(k)$,
which is called the dispersion relation, after substituting this solution 
into (\ref{eq:dynsurflin}).

When the dissipative term is chosen according to model I (\ref{eq:modelI}), 
$D_\phi = \delta_1\phi$ and the dispersion relation is given implicitly by
\begin{equation*}
  \omega^2 + i\delta_1\omega -gk\tanh(kh) = 0,
\end{equation*}
or in explicit form by
\begin{equation}\label{eq:disprelI}
  \omega = \pm\sqrt{gk\tanh(kh) - \frac{\delta_1^2}4} - \frac{i\delta_1}{2}.
\end{equation}

For the second dissipation model (\ref{eq:modelII}) one obtains the following
relation:
\begin{equation*}
  \omega^2 + i\delta_2\omega k^2 - gk\tanh(kh) = 0.
\end{equation*}
One can easily solve this quadratic equation for
$\omega$ as a function of $k$:
\begin{equation}\label{eq:disprelII}
  \omega = \pm\sqrt{gk\tanh(kh) - \left(\frac{\delta_2 k^2}{2}\right)^2} - \frac{i\delta_2}{2}k^2.
\end{equation}

If $\delta_{1,2}\equiv 0$ one easily recognizes the dispersion relation of the classical water-wave problem: 
\begin{equation}\label{eq:fulldisprel}
	\omega = \pm\sqrt{gk\tanh(kh)}.
\end{equation}
\textbf{Remark:} It is important to have the property $\Im\omega(k) \leq 0, \forall k$  in order to avoid 
the exponential growth of certain wavelengths, since
\begin{equation*}
  e^{i(kx-\omega(k) t)} = e^{\Im \omega(k) t}\cdot e^{i(kx-\Re\omega(k) t)}.
\end{equation*}

For our analysis it is more interesting to look at the phase speed which is defined as
\begin{equation*}
  c_{p}(k) := \frac{\omega(k)}{k}.
\end{equation*}
The phase velocity is directly connected to the speed of wave propagation and is extremely important 
for accurate tsunami modelling since tsunami arrival time obviously depends on the propagation 
speed. The expressions for the phase velocity are obtained from the corresponding
dispersion relations (\ref{eq:disprelI}) and (\ref{eq:disprelII}):
\begin{equation}\label{eq:disprelfull1}
  c_p^{(1)}(k) = \pm\sqrt{gh\frac{\tanh(kh)}{kh} - \Bigl(\frac{\delta_1}{2k}\Bigr)^2} - \frac{i\delta_1}{2k},
\end{equation}
\begin{equation}\label{eq:disprelfull2}
  c_p^{(2)}(k) = \pm\sqrt{gh\frac{\tanh(kh)}{kh} - \Bigl(\frac{\delta_2 k}{2}\Bigr)^2} - \frac{i\delta_2}{2}k.
\end{equation}
It can be shown that in order to keep the phase velocity unchanged by the addition of dissipation, similar dissipative
terms must be included in both the kinematic and the dynamic boundary conditions \citep{ddz}.

\subsection{Dissipative Boussinesq equations}

The analysis of the dispersion relation is even more straightforward for Boussinesq equations.
In order to be coherent with the previous subsection, we switch to dimensional 
variables. As usual we begin with the $(1+1)$D linearized equations:
\begin{equation*}
  \eta_t + h u_x = \frac{h^3}{6}u_{xxx},
\end{equation*}
\begin{eqnarray*}
  \mbox{ Model I: } & u_t + g\eta_x + \delta_1 u = \frac12\delta_1h^2 u_{xx} + \frac12{h^2}u_{xxt}, \\
  \mbox{Model II: } & u_t + g\eta_x           = \delta_2 u_{xx} + \frac12{h^2}u_{xxt}.
\end{eqnarray*}
Now we substitute a special ansatz in these equations:
\begin{equation*}
  \eta = \eta_0 e^{i(kx-\omega t)}, \qquad
  u = u_0 e^{i(kx-\omega t)},
\end{equation*}
where $\eta_0$ and $u_0$ are constants. In the case of the first model,
one obtains the following homogeneous system of linear equations:
\begin{eqnarray*}
  (-i\omega)\eta_0 + ikh\left(1 + \frac16(kh)^2\right)u_0 & = 0, \\
  g ik\eta_0 + \left(-i\omega + \delta_1 + \frac{\delta_1}{2}(kh)^2 - \frac{i\omega}2 (kh)^2\right)u_0 & = 0.
\end{eqnarray*}
This system admits nontrivial solutions if its determinant is equal to zero. It gives the
required dispersion relation:
\begin{equation*}
  \omega^2 + i\omega\delta_1 - gh k^2\left(\frac{1+\frac16 (kh)^2}{1+\frac12 (kh)^2}\right) = 0.
\end{equation*}
A similar relation is found for the second model:
\begin{equation*}
  \omega^2 + \frac{i\omega\delta_2}{1+\frac12(kh)^2} k^2 - gh k^2\left(\frac{1+\frac16(kh)^2}{1+\frac12(kh)^2}\right) = 0.
\end{equation*}
The corresponding phase velocities are given by 
\begin{equation}\label{eq:disprelbouss1}
  c_{pb}^{(1)} = \sqrt{gh\left(\frac{1+\frac16(kh)^2}{1+\frac12(kh)^2}\right) - \left(\frac{\delta_1}{2k}\right)^2} 
  - \frac{i\delta_1}{2k},
\end{equation}
\begin{equation}\label{eq:disprelbouss2}
  c_{pb}^{(2)} = \sqrt{gh\left(\frac{1+\frac16(kh)^2}{1+\frac12(kh)^2}\right) - \left(\frac{\delta_2k}{2+(kh)^2}\right)^2} 
  - \frac{i\delta_2k}{2+(kh)^2}.
\end{equation}

\begin{figure}
	\centering
		\includegraphics[width=0.90\textwidth]{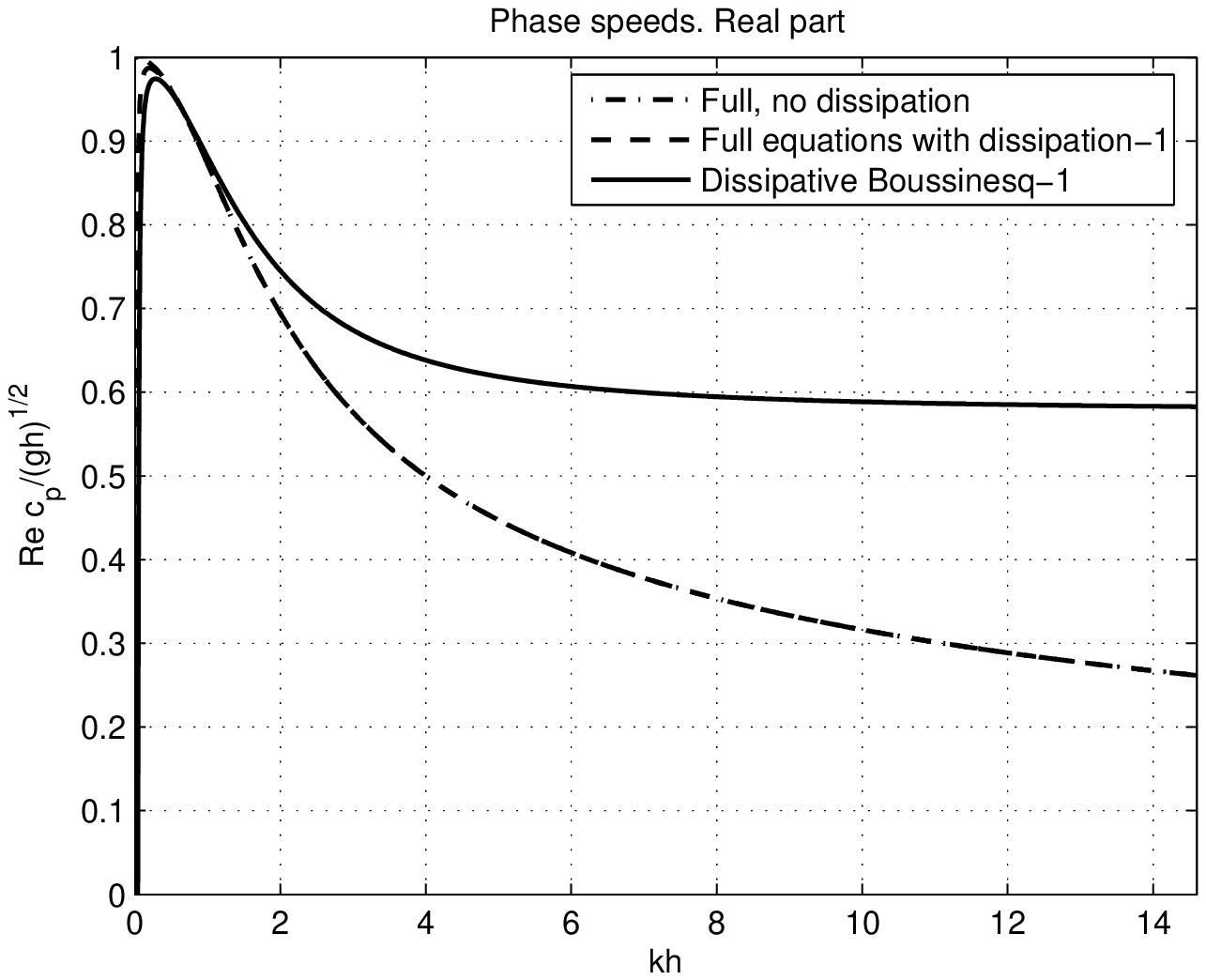}
	\caption{Dissipation model I. Real part of the phase velocity.}
	\label{fig:model1real}
\end{figure}

\begin{figure}
	\centering
		\includegraphics[width=0.90\textwidth]{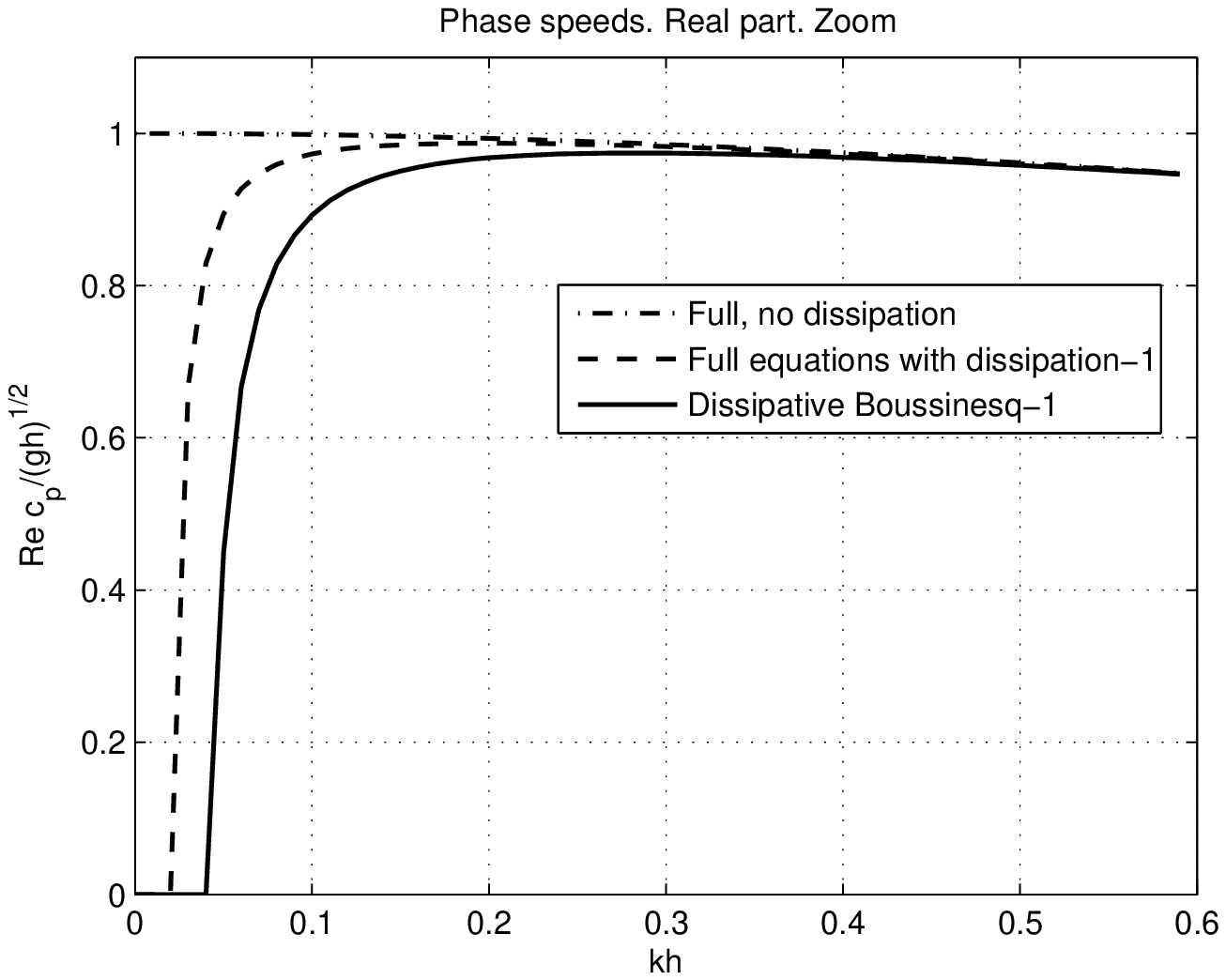}
	\caption{Dissipation model I. Same as Figure \ref{fig:model1real} with a zoom on long waves.}
	\label{fig:model1realzoom}
\end{figure}

\begin{figure}
	\centering
		\includegraphics[width=0.90\textwidth]{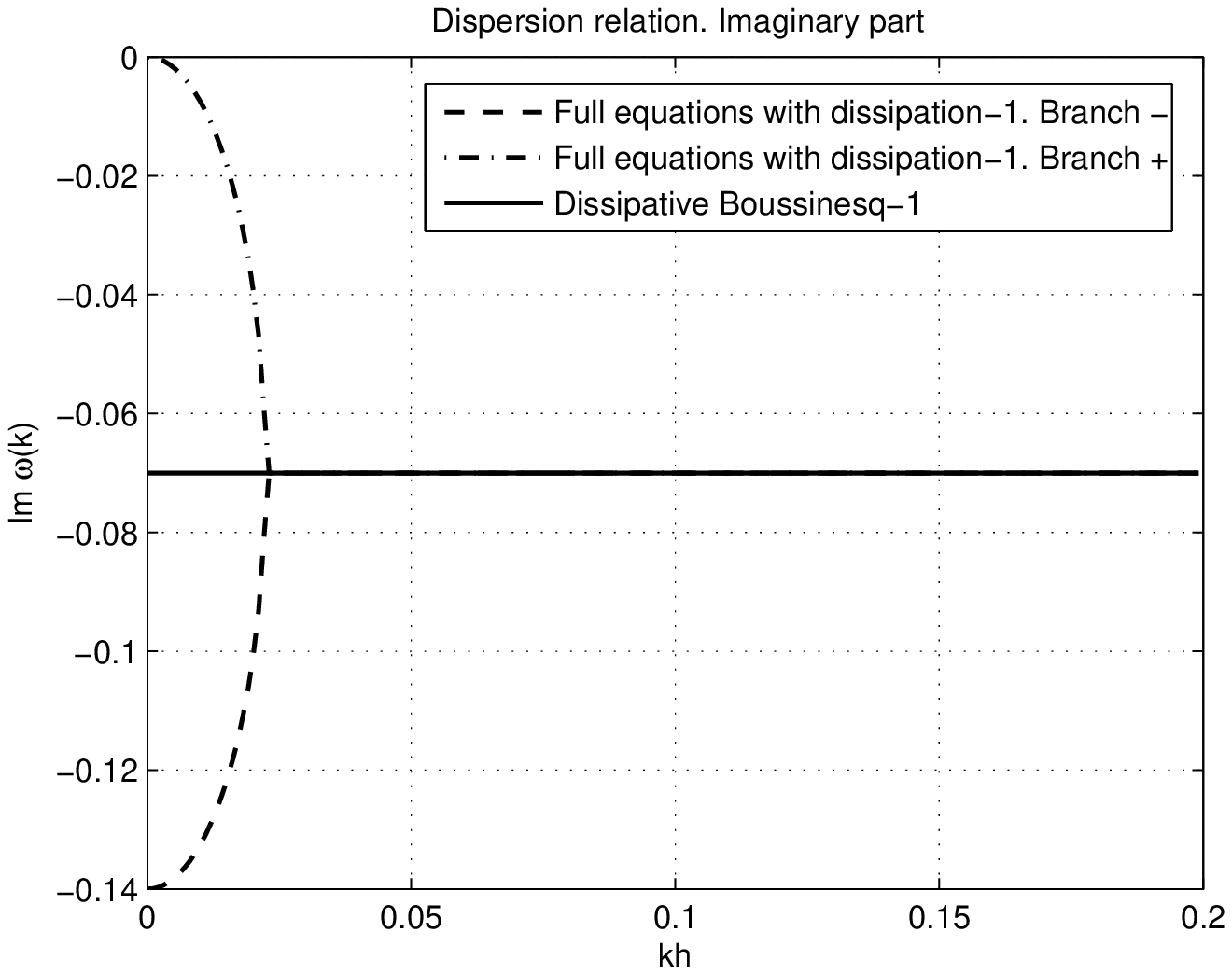}
	\caption{Dissipation model I. Imaginary part of the frequency.}
	\label{fig:model1imag}
\end{figure}

\begin{figure}
	\centering
		\includegraphics[width=0.90\textwidth]{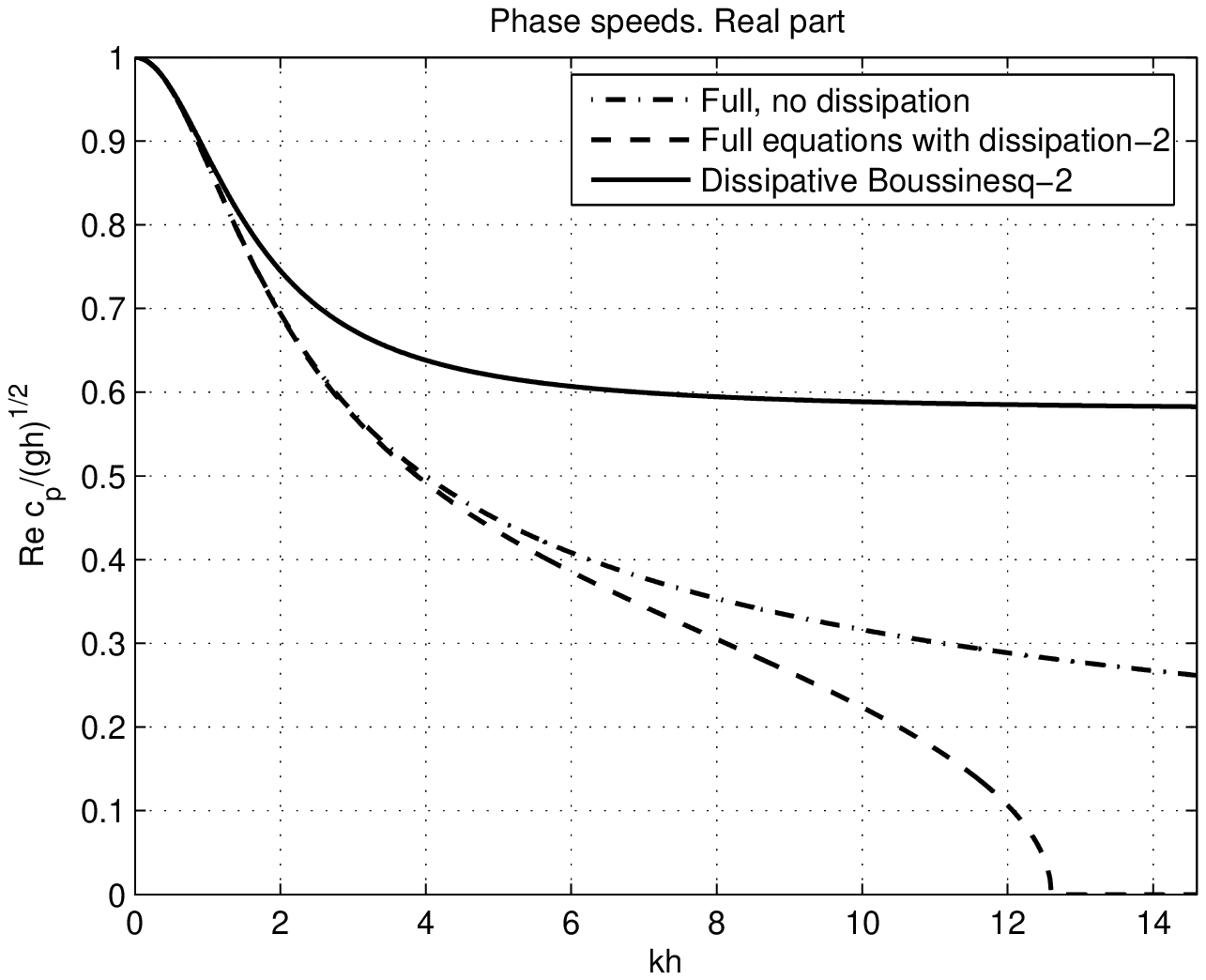}
	\caption{Dissipation model II. Real part of the phase velocity.}
	\label{fig:model2real}
\end{figure}

\begin{figure}
	\centering
		\includegraphics[width=0.85\textwidth]{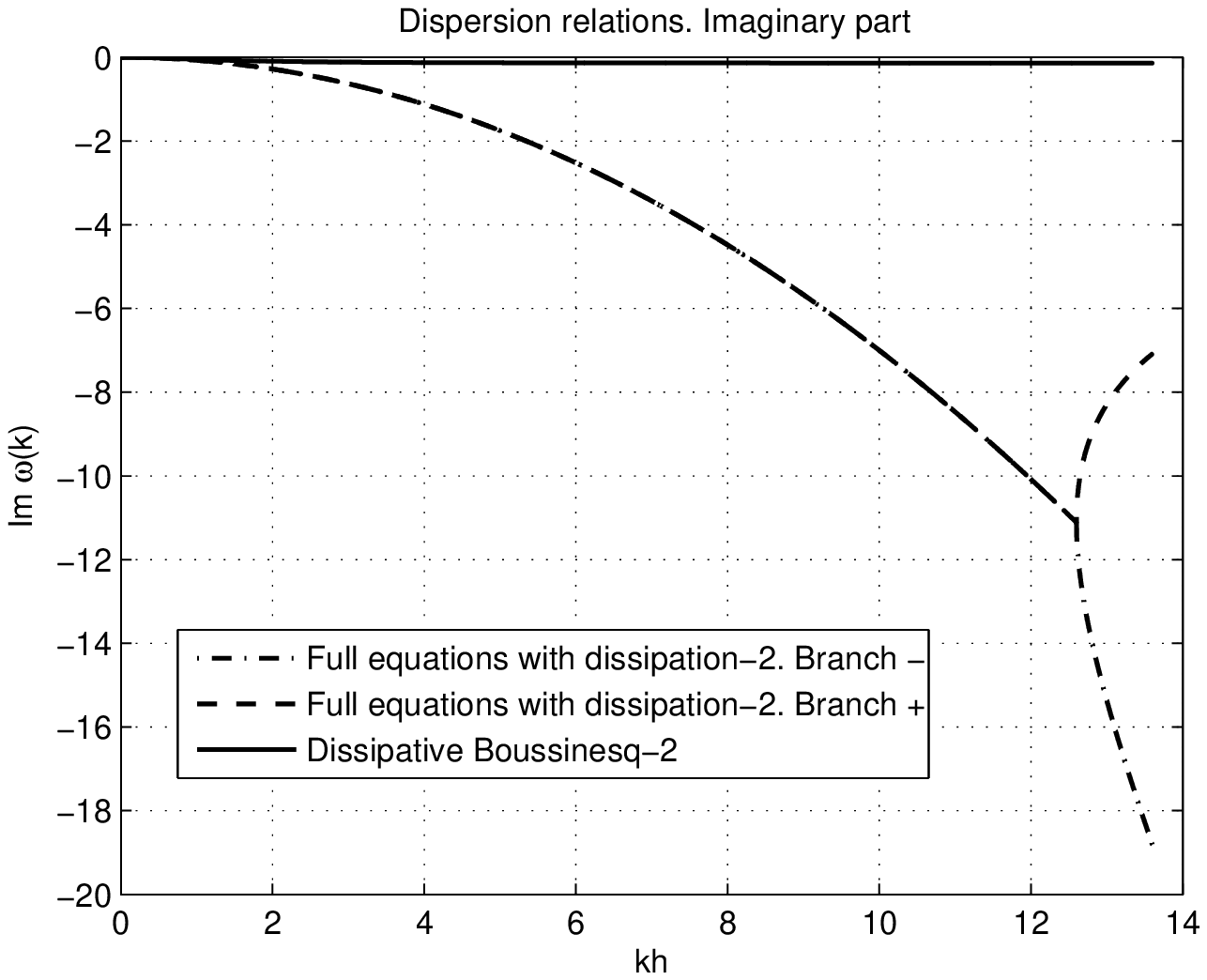}
	\caption{Dissipation model II. Imaginary part of the frequency.}
	\label{fig:model2imag}
\end{figure}

\begin{figure}
	\centering
		\includegraphics[width=0.90\textwidth]{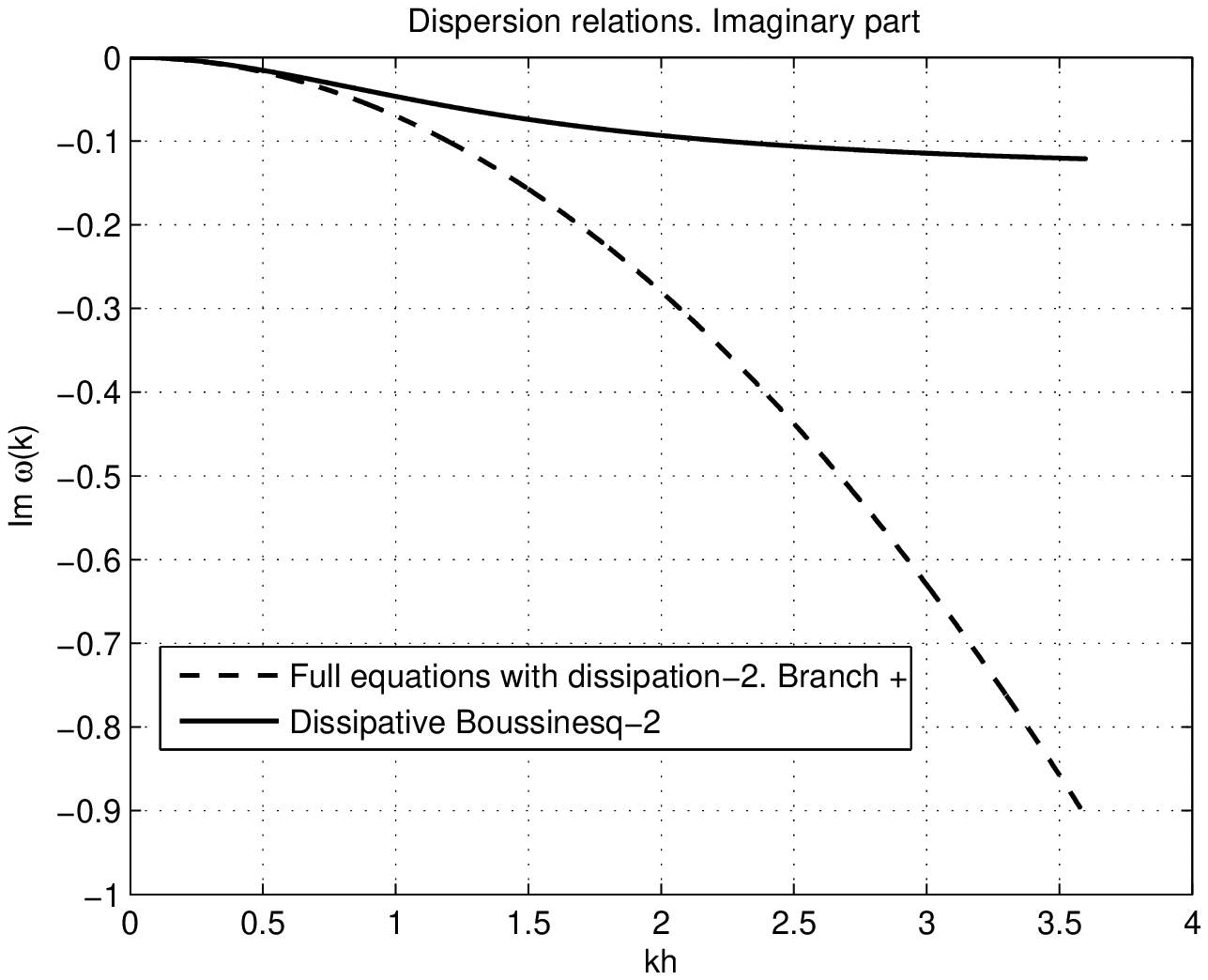}
	\caption{Dissipation model II. Same as Figure \ref{fig:model2imag} with a zoom on long waves.}
	\label{fig:model2ImagKzoom}
\end{figure}

\subsection{Discussion}

Let us now provide a discussion on the dispersion relations. The real and imaginary parts 
of the phase velocities (\ref{eq:disprelfull1})--(\ref{eq:disprelbouss2}) 
for the full and long wave linearized equations are shown graphically on 
Figures \ref{fig:model1real}--\ref{fig:model2ImagKzoom}. In this example the parameters are given by
$\delta_1 = 0.14, \delta_2 = 0.14$.
Together with the dissipative models we also plotted for comparison the well-known
phase velocity corresponding to the full conservative (linearized) water-wave problem:
\begin{equation*}
  c_p(k) = \sqrt{gh\frac{\tanh(kh)}{kh}}.
\end{equation*}

First of all, one can see that
dissipation is very selective, as is often the case in physics. Clearly, the first dissipation
model prefers very long waves, while the second model dissipates essentially short
waves. Moreover one can see from the expressions (\ref{eq:disprelfull1}),
(\ref{eq:disprelbouss1}) that the phase velocity has a $1/k$ singular behaviour
in the vicinity of $kh=0$ (in the long wave limit). Furthermore, it can be clearly
seen in Figure \ref{fig:model1realzoom} that very long linear waves are not advected
in the first dissipation model, since the real part of their phase velocity is identically equal
to zero.

That is why we suggest to make use of the second model in applications involving 
very long waves such as tsunamis.

On the other hand we would like to point out that the second model admits a
critical wavenumber $k_c$ such that the phase velocity (\ref{eq:disprelfull2})
becomes purely imaginary with negative imaginary part. From a physical point of
view it means that the waves shorter than $k_c$ are not advected, but
only dissipated. When one switches to the Boussinesq approximation, this property disappears
for physically realistic values of the parameters $g$, $h$ and $\delta_2$ (see Table \ref{tab:params}).

Let us clarify this situation. The qualitative behaviour of the phase velocity
$c_{pb}^{(2)}$ (see equation (\ref{eq:disprelbouss2})) depends on the roots 
of the following polynomial equation:
\begin{equation*}
  (kh)^4 + \left(8 - \frac{3\delta_2^2}{gh^3}\right)(kh)^2 + 12 = 0.
\end{equation*}
This equation does not have real roots since $\frac{3\delta_2^2}{gh^3}\ll 1$.

\section{Alternative version of the Boussinesq equations}

In this section we give an alternative derivation of Boussinesq equations.
We use another classical method for deriving Boussinesq-type equations 
\citep{Whitham1999, Benjamin1974, Peregrine1972},
which provides slightly different governing equations. Namely, the hyperbolic 
structure is the same, but the dispersive terms differ. In numerical simulations we
suggest to use this system of equations.

The derivation follows closely the paper by \cite{Madsen1998}. 
The main differences are that we neglect the terms of order $O(\mm)$, take
in account a moving bathymetry and, of course, dissipative effects which
are modelled this time according to model II (\ref{eq:modelII}) because,
in our opinion, this model is more appropriate for long wave applications.
Anyhow, the derivation process can be performed in a similar fashion
for model I (\ref{eq:modelI}).

\subsection{Derivation of the equations}

The starting point is the same: equations (\ref{eq:conteq}), (\ref{eq:freesurfkin}),
(\ref{eq:dynbound}) and (\ref{eq:kinembound}). This time the procedure begins with
representing the velocity potential $\phi(x,y,z,t)$ as a formal expansion
in powers of $z$ rather than of $\m$:
\begin{equation}\label{eq:zexpansion}
  \phi(x,y,z,t) = \sum_{n=0}^{\infty} z^n \phi_n(x,y,t).
\end{equation}
We would like to emphasize that this expansion is only formal and no convergence
result is needed. In other words, it is just convenient to use this notation
in asymptotic expansions but in practice, seldom more than four terms are used.
It is not necessary to justify the convergence of the sum with 
three or four terms.

When we substitute the expansion (\ref{eq:zexpansion}) into Laplace equation (\ref{eq:conteq}),
we have an infinite polynomial in $z$. Requiring that $\phi$ formally satisfies
Laplace equation implies that the coefficients of each power of $z$
vanish (since the right-hand side is identically zero). This leads to the classical
recurrence relation
\begin{equation*}
  \phi_{n+2}(x,y,t) = -\frac{\m}{(n+1)(n+2)}\nabla^2\phi_n(x,y,t), \qquad
  n = 0, 1, 2, \ldots
\end{equation*}
Using this relation one can eliminate all but two unknown functions in (\ref{eq:zexpansion}):
\begin{equation*}
  \phi(x,y,z,t) = \sum_{n=0}^{\infty} (-1)^n\mu^{2n} \left(
  \frac{z^{2n}}{(2n)!}\nabla^{2n}\phi_0 + \frac{z^{2n+1}}{(2n+1)!}\nabla^{2n}\phi_1
  \right).
\end{equation*}
The following notation is introduced:
\begin{equation*}
  \u_0 := \u(x,y,0,t), \qquad
  w_0 := \frac{1}{\m} w(x,y,0,t).
\end{equation*}
It is straightforward to find the relations between $\u_0$, $w_0$ and $\phi_0$, $\phi_1$
if one remembers that $(\u,w) = (\nabla,\pd{}{z})\phi$:
\begin{equation*}
  \u_0 = \nabla \phi_0, \qquad
  w_0 = \frac{1}{\m}\phi_1.
\end{equation*}
Using the definition of the velocity potential $\phi$
one can express the velocity field in terms of $\u_0$, $w_0$:
\begin{equation*}
 \u = \sum_{n=0}^{\infty} (-1)^n 
 \left(\frac{z^{2n}}{(2n)!}\mu^{2n}\nabla\bigl(\nabla^{2n-2}(\nabla\cdot\u_0)\bigr)
 +\frac{z^{2n+1}}{(2n+1)!}\mu^{2n+2}\nabla\bigl(\nabla^{2n}w_0\bigr)\right),
\end{equation*}
\begin{equation*}
  w = \sum_{n=0}^{\infty} (-1)^n\left(
  -\frac{z^{2n+1}}{(2n+1)!} \mu^{2n+2}\nabla^{2n}(\nabla\cdot\u_0) + 
  \frac{z^{2n}}{(2n)!}\mu^{2n+2}\nabla^{2n}w_0\right).
\end{equation*}
These formulas are exact but not practical. In the present work we 
neglect the terms of order $O(\mm)$ and higher. In this asymptotic 
framework the above formulas become much simpler:
\begin{equation}\label{eq:phiexp2}
  \phi = \phi_0 + z\phi_1 - \frac{\m z^2}{2}\left(\nabla^2\phi_0 + 
  \frac{z}{3}\nabla^2\phi_1\right) + O(\mm),
\end{equation}
\begin{equation}\label{eq:uexp}
  \u = \u_0 + z\m\nabla w_0 - \frac{\m z^2}{2}\nabla(\nabla\cdot\u_0) + O(\mm),
\end{equation}
\begin{equation}\label{eq:wexp}
  w = \m w_0 - z\m\nabla\cdot \u_0 + O(\mm).
\end{equation}

In order to establish the relation between $w_0$ and $\u_0$ one uses the bottom kinematic 
boundary condition (\ref{eq:kinembound}), which has the following form 
after substituting the asymptotic expansions (\ref{eq:phiexp2}), (\ref{eq:uexp}),
(\ref{eq:wexp}) in it:
\begin{multline}\label{eq:bottomexpansion}
  h_t + \eps\nabla h\cdot\Bigl(\u_0 - h\m\nabla w_0 - 
  \frac{\m h^2}{2}\nabla(\nabla\cdot\u_0)\Bigr) \\ + \eps\Bigl(w_0 - 
  \frac{h^3}{6}\m\nabla^2(\nabla\cdot\u_0) - \frac{\m h^2}{2}\nabla^2 w_0\Bigr) + O(\mm) = 0.
\end{multline}

In order to obtain the expression of $w_0$ in terms of $\u_0$ one introduces one
more expansion:
\begin{equation}\label{eq:wexpansion}
  w_0(x,y,t)  = w_0^{(0)}(x,y,t) + \m w_0^{(1)}(x,y,t) + \ldots.
\end{equation}
We insert this expansion into the asymptotic bottom boundary condition (\ref{eq:bottomexpansion}).
This leads to the following explicit expressions for $w_0^{(0)}$ and $w_0^{(1)}$:
\begin{equation*}
  w_0^{(0)} = -\frac{1}{\eps}h_t - \nabla\cdot(h\u_0),
\end{equation*}
\begin{multline*}
  w_0^{(1)} = \frac{h^2}{2}\Bigl(\nabla h\cdot\nabla(\nabla\cdot\u_0) + 
  \frac{h}{3}\nabla^2(\nabla\cdot\u_0)\Bigr) \\ - 
  h\left(\nabla h\cdot\frac{1}{\eps}\nabla h_t + 
  \nabla h\cdot\nabla\bigl(\nabla\cdot(h\u_0)\bigr) + 
  \frac{h}{2}\bigl(\frac{1}{\eps}\nabla^2h_t+\nabla^2(\nabla\cdot(h\u_0))\bigr)\right).
\end{multline*}
Substituting these expansions into (\ref{eq:wexpansion}) and performing some 
simplifications yields the required relation between $\u_0$ and $w_0$:
\begin{multline}\label{eq:verticalvelocity}
  w_0 = -\frac1{\eps}h_t - \nabla\cdot(h\u_0) \\- 
  \m\nabla\cdot\Bigl(\frac{h^2}{2\eps}\nabla h_t + 
  \frac{h^2}{2}\nabla\bigl(\nabla\cdot(h\u_0)\bigr)
  -\frac{h^3}{6}\nabla(\nabla\cdot\u_0)\Bigr) + O(\mm).
\end{multline}
Now one can eliminate the vertical velocity $w_0$ since one has its expression 
(\ref{eq:verticalvelocity}) in terms of $\u_0$. Equations (\ref{eq:uexp})-(\ref{eq:wexp}) become
\begin{equation}\label{eq:horvelocity}
  \u = \u_0 - z\frac{\m}{\eps}\nabla h_t - 
  \m\Bigl(z\nabla\bigl(\nabla\cdot(h\u_0)\bigr) + \frac{z^2}{2}\nabla(\nabla\cdot\u_0)\Bigr)
  + O(\mm),
\end{equation}
\begin{equation}\label{eq:vertvelocity}
  w = -\frac{\m}{\eps}h_t - 
  \m\Bigl(\nabla\cdot(h\u_0) + z\nabla\cdot\u_0\Bigr) + O(\mm).
\end{equation}

In this work we apply a trick due to \cite{Nwogu1993}. Namely, we introduce a new velocity variable 
$\ua$ defined at an arbitrary water level $\za=-\alpha h$.
Technically this change of variables is done as follows. First we evaluate
(\ref{eq:horvelocity}) at $z=\za$, which gives the connection between $\u_0$ and $\ua$:
\begin{equation*}
  \ua = \u_0 - \za\frac{\m}{\eps}\nabla h_t -
  \m\Bigl(\za\nabla\bigl(\nabla\cdot(h\u_0)\bigr) + \frac{\za^2}{2}\nabla(\nabla\cdot\u_0)\Bigr)
  + O(\mm).
\end{equation*}
Using the standard technics of inversion one can rewrite the last expression as an asymptotic 
formula for $\u_0$ in terms of $\ua$:
\begin{equation}\label{eq:u0ua}
  \u_0 = \ua + \za\frac{\m}{\eps}h_t + 
  \m\Bigl(\za\nabla\bigl(\div(h\ua)\bigr) + \frac{\za^2}{2}\nabla(\div\ua)\Bigr) + O(\mm).
\end{equation}

\textbf{Remark:} Behind this change of variables there is one subtlety which is
generally hushed up in the literature. In fact, the wave motion is assumed
to be irrotational since we use the potential flow formulation (\ref{eq:conteq}), (\ref{eq:freesurfkin}),
(\ref{eq:dynbound}), (\ref{eq:kinembound}) of the water-wave problem. 
By construction $\rot(\u,w) = \O$ when $\u$ and $w$ are computed according to (\ref{eq:horvelocity}),
(\ref{eq:vertvelocity}) or, in other words, in terms of the variable $\u_0$. When one turns to the
velocity variable $\ua$ defined at an arbitrary level, one can improve the linear dispersion relation
and this is important for wave modelling. But on the other hand, one loses the property 
that the flow is irrotational. That is to say, a direct computation shows that 
$\rot(\u,w)\neq\O$ when $\u$ and $w$ are expressed in terms of the variable $\ua$. 
The purpose of this remark is simply to inform the reader about the price to be paid
 while improving the dispersion relation properties. It seems that
this point is not clearly mentioned in the literature on this topic.

Let us now derive the Boussinesq equations. There are two different methods to obtain
the free-surface elevation equation. 
The first method consists in integrating the continuity equation (\ref{eq:conteq})
over the depth and then use the kinematic free-surface and bottom boundary conditions.
The second way is more straightforward. It consists in using directly 
the kinematic free-surface boundary condition (\ref{eq:freesurfkin}):
\begin{equation*}
	\eta_t + \eps\nabla\phi\cdot\nabla\eta -\frac{1}{\m}\phi_z = 0.
\end{equation*}
Then one can substitute (\ref{eq:phiexp2}) into (\ref{eq:freesurfkin}) and perform several 
simplifications. Neglecting all terms of order $O(\eps^2+\eps\m+\mm)$ 
yields the following equation\footnote{We already discussed this 
point on page \pageref{page:Stokes}. In this section we also assume that the Stokes-Ursell
number $S$ is of order $O(1)$.} :
\begin{multline*}
	\eta_t + \div\bigl((h+\eps\eta)\u_0\bigr) + 
	\frac{\m}{2}\div\Bigl(h^2\nabla\bigl(\div(h\u_0)\bigr)
	-\frac{h^3}{3}\nabla(\div\u_0)\Bigr) = \\ =
	\zeta_t + \frac{\m}{2}\div(h^2\nabla\zeta_t).
\end{multline*}
Recall that $\zeta(x,y,t)$ is defined according to (\ref{eq:defzeta}).
When the bathymetry is static, $\zeta\equiv 0$. We prefer to introduce this
function in order to eliminate the division by $\eps$ in the source terms since
this division can give the impression that stiff source terms are present in our problem,
which is not the case.

In order to be able to optimize the dispersion relation properties, we switch to the
variable $\ua$. Technically it is done by using the relation (\ref{eq:u0ua}) between 
$\u_0$ and $\ua$. The result is given below:
\begin{multline}\label{eq:main_eta}
  \eta_t + \div\bigl((h+\eps\eta)\ua\bigr) 
  + \m\div\Bigl(h\bigl(\za+\frac{h}2\bigr)\nabla\bigl(\div(h\ua)\bigr) 
  + \frac{h}{2}\bigl(\za^2-\frac{h^2}{3}\bigr)\nabla(\div\ua)\Bigr) 
  =\\= \zeta_t + \m\div\Bigl(h\bigl(\za+\frac{h}2\bigr)\nabla\zeta_t\Bigr).
\end{multline}

As above, the equation for the horizontal velocity field is derived from the dynamic
free-surface boundary condition (\ref{eq:dynbound}). It is done exactly as in section \ref{sec:section1} and 
we do not insist on this point:
\begin{equation*}
	\u_{0t} + \frac{\eps}{2}\nabla\abs{\u_0}^2 + \nabla\eta 
	- \eps\delta\nabla^2\u_0 = \O.
\end{equation*}
Switching to the variable $\ua$ yields the following governing equation:
\begin{multline}\label{eq:main_u}
  \ua_t + \frac{\eps}{2}\nabla|\ua|^2 + \nabla\eta
   + \m\Bigl(\za\nabla\bigl(\div(h\ua)\bigr) + \frac{\za^2}{2}\nabla(\div\ua)\Bigr)_t =\\=
   \eps\delta\Delta\ua + \m(\za\nabla\zeta_t)_t.
\end{multline}

In several numerical methods it can be advantageous to rewrite the system
(\ref{eq:main_eta}), (\ref{eq:main_u}) in vector form:
\begin{equation*}
 \U_t + \m\L(\U)_t + \div \F(\U) + \m\div \P(\U) = \S(x,y,t) + \eps\delta\div(\D\nabla\U)),
\end{equation*}
where
\begin{equation*}
  \U := 
  \begin{pmatrix}
    \eta \\
    u_\alpha \\
    v_\alpha
  \end{pmatrix}, \quad
  \div\F := \pd{F}{x} + \pd{G}{y},
\end{equation*}

\begin{equation*}
  F := \begin{pmatrix}
    (h+\eps\eta)u_\alpha \\
    \frac{\eps}{2}|\ua|^2 + \eta \\
    0
  \end{pmatrix}, \quad
  G := \begin{pmatrix}
    (h+\eps\eta) v_\alpha \\
    0 \\
    \frac{\eps}{2}|\ua|^2 + \eta \\
  \end{pmatrix},
\end{equation*}

\begin{equation*}
  \L := \begin{pmatrix}
    0 \\
    \za\nabla\bigl(\div(h\ua)\bigr) + \frac{\za^2}{2}\nabla(\div\ua)
  \end{pmatrix},
\end{equation*}

\begin{equation*}
  \P := \begin{pmatrix}
    h\bigl(\za+\frac{h}2\bigr)\nabla\bigl(\div(h\ua)\bigr) 
  + \frac{h}{2}\bigl(\za^2-\frac{h^2}{3}\bigr)\nabla(\div\ua) \\
    \O
  \end{pmatrix},
\end{equation*}

\begin{equation*}
  \S := \begin{pmatrix}
    \zeta_t + \m\div\Bigl(h\bigl(\za+\frac{h}2\bigr)\nabla\zeta_t\Bigr) \\
    \m(\za\nabla\zeta_t)_t
  \end{pmatrix},
\end{equation*}

\begin{equation*}
 \D := \begin{pmatrix}
 0 & 0 & 0 \\
 0 & 1 & 0 \\
 0 & 0 & 1 \\
 \end{pmatrix}.
\end{equation*} 

\subsection{Improvement of the linear dispersion relations}

As said above, the idea of using one free parameter $\alpha\in[0,1]$
to optimize the linear dispersion relation properties appears to have been proposed
first by \cite{Nwogu1993}.

The idea of manipulating the dispersion relation was well-known before 1993.
See for example \cite{Murray1989, Madsen1991}. But
these authors started with a desired dispersion relation and artificially added
extra terms to the momentum equation in order to produce the desired characteristics.
We prefer to follow the ideas of \cite{Nwogu1993}.

\textbf{Remark:} When one plays with the dispersion relation it is important to
remember that the resulting problem must be well-posed, at least linearly.
We refer to \cite{BCS} as a general reference on this topic. Usually Boussinesq-type models with good dispersion characteristics
are linearly well-posed as well.

In order to look for an optimal value of $\alpha$ we will
drop dissipative terms. Indeed we want to
concentrate our attention on the propagation properties which are more important.

The choice for the parameter $\alpha$ depends on the optimization criterion. 
In the present work we
choose $\alpha$ by comparing the coefficients in the Taylor expansions of 
the phase velocity in the vicinity of $kh=0$, which corresponds to the long-wave limit.
Another possibility is to match the dispersion relation of the full linearized equations 
(\ref{eq:fulldisprel}) in the least square sense. One can also use Padé approximants \citep{Witting1984} since rational
functions have better approximation properties than polynomials. 

We briefly describe the procedure. First of all one has to obtain 
the phase velocity of the linearized, non-viscous, Boussinesq equations
(\ref{eq:main_eta})-(\ref{eq:main_u}). The result is
\begin{equation}\label{eq:disprelalpha}
	\frac{c_b^2(k)}{gh} = 
	\frac{1-\bigl(\frac{\alpha^2}{2}-\alpha+\frac13\bigr)(kh)^2}%
	{1-\alpha\bigl(\frac{\alpha}{2}-1\bigr)(kh)^2} = 
	1 - \frac13(kh)^2 + \frac{\alpha(2-\alpha)}{6}(kh)^4 + O\left((kh)^6\right).
\end{equation}
On the other hand one can write down the phase velocity of the full linearized equations 
(\ref{eq:fulldisprel}):
\begin{equation*}
	\frac{c^2(k)}{gh} = \frac{\tanh(kh)}{kh} =
	1 - \frac13(kh)^2 + \frac{2}{15}(kh)^4 + O\left((kh)^6\right).
\end{equation*}
If one insists on the dispersion relation (\ref{eq:disprelalpha}) to be exact 
up to order $O\left((kh)^4\right)$ one immediately obtains an equation for $\alpha_{opt}$:
\begin{equation*}
	\frac{\alpha_{opt}(2-\alpha_{opt})}{6} = \frac{2}{15} \Rightarrow
	\alpha_{opt} = 1 - \frac{\sqrt{5}}{5} \approx 0.55.
\end{equation*}
We suggest using this value of $\alpha$ in numerical computations.

\subsection{Bottom friction}

In this subsection, one switches back to dimensional variables. 
It is a common practice in hydraulics engineering to take into account the effect
of bottom friction or bottom rugosity. In the Boussinesq and nonlinear shallow water equations there is also a possibility to
include some kind of empirical terms to model these physical effects.
From the mathematical and especially numerical viewpoints these
terms do not add any complexity, since they have the form of source terms that do 
not involve differential operators. So it is highly recommended to introduce 
these source terms in numerical models. 

There is no unique bottom friction law. Most frequently, 
Chézy and Darcy-Weisbach laws are used. Both laws have similar structures.
We give here these models in dimensional form. 
The following terms have to be added to the source terms of Boussinesq equations
when one wants to include bottom friction modelling.

\begin{itemize}
\item Chézy law:
\begin{equation*}
 \S_f = -C_fg\frac{\u\abs{\u}}{h+\eta},
\end{equation*}
where $C_f$ is the Chézy coefficient.
\item Darcy-Weisbach law:
\begin{equation*}
 \S_f = -\frac{\lambda\u\abs{\u}}{8(h+\eta)},
\end{equation*}
where $\lambda$ is the so-called resistance value. This parameter is determined
according to the simplified form of the Colebrook-White relation:
\begin{equation*}
 \frac{1}{\sqrt{\lambda}} = -2.03 \log\left(\frac{k_s}{14.84(h+\eta)}\right).
\end{equation*}
Here $k_s$ denotes the friction parameter, which depends on the composition
of the bottom. Typically $k_s$ can vary from $1$mm for concrete to $300$mm
for bottom with dense vegetation.
\item Manning-Strickler law:
\begin{equation*}
 \S_f = -k^2 g \frac{\u\abs{\u}}{(h+\eta)^{\frac43}},
\end{equation*}
where $k$ is the Manning roughness coefficient.
\end{itemize} 

\section{Spectral Fourier method}

In this study we adopted a well-known and widely used spectral Fourier method. 
The main idea consists in
discretizing the spatial derivatives using Fourier transforms. The effectiveness
of this method is explained by two main reasons. First, the differentiation
operation in Fourier transform space is extremely simple due to the following 
property of Fourier transforms: $\overline{f'} = ik \overline{f}$.
Secondly, there are very powerful tools for the fast
and accurate computation of discrete Fourier transforms (DFT). 
So, spatial derivatives are computed with the following algorithm:
\begin{algorithmic}[1]
	\STATE $\overline f \leftarrow \fft(f)$
	\STATE $\overline v \leftarrow ik\overline f$
	\STATE $f' \leftarrow \ifft(\overline v)$
\end{algorithmic}
where $k$ is the wavenumber.

This approach, which is extremely efficient, has the drawbacks of almost all spectral methods. The first drawback consists in
imposing periodic boundary conditions since we use DFT. The second drawback is
that we can only handle simple geometries, namely, Cartesian products
of $1$D intervals. For the purpose
of academic research, this type of method is appropriate.

Let us now consider the discretization of the dissipative 
Boussinesq equations. We show in detail how the discretization is performed
on equations (\ref{eq:etacommon}), (\ref{eq:umodelII}).
The other systems are discretized in the same way. 
We chose equations (\ref{eq:etacommon}), (\ref{eq:umodelII}) in order to
avoid cumbersome expressions and make the description as clear as possible.

Let us apply the Fourier transform to both sides of equations (\ref{eq:etacommon}), 
(\ref{eq:umodelII}):
\begin{multline}\label{eq:etadiff}
 \overline\eta_t = -i\k\cdot\overline{(h+\eps\eta)\u} - \frac1\eps\overline h_t -
 \frac{\m}{2\eps}\overline{h^2\nabla^2h_t} -
 \frac\m\eps\overline{h\nabla h\cdot\nabla h_t} +
 \frac{\m}{6}\overline{h^3\nabla^2\div\u} \\ +
 \frac{\m}{2}\overline{h^2\nabla^2h\div\u} + \m\overline{h\abs{\nabla h}^2\div\u} +
 \m\overline{h^2\nabla h\cdot\nabla(\div\u)},
\end{multline}
\begin{equation}\label{eq:ubardiff}
 \overline\u_t + \frac12 {\eps}i\k\overline{\abs{\u}^2} + i\k\overline\eta +
 \eps\nu_2\abs{\k}^2\overline\u -\frac12 \m i\k\overline{h^2\div\u_t} = 0,
\end{equation}
where $\k=(k_x,k_y)$ denotes the Fourier transform parameters.

Equations (\ref{eq:etadiff}) and (\ref{eq:ubardiff}) constitute a system
of ordinary differential equations to be integrated numerically. In the present
study we use the classical explicit fourth-order Runge-Kutta method.

\textbf{Remark on stability:} A lot of researchers
who integrated numerically the KdV equation noticed that the stability criterion
has the form
\begin{equation*}
	\Delta t = \frac{\lambda}{N^2},
\end{equation*}
where $\lambda$ is the Courant-Friedrichs-Lewy (CFL) number and $N$ the number of points of discretization. 
In order to increase the time integration
step $\Delta t$ they solved exactly the linear part of the partial differential equation since the linear term is 
the one involving high frequencies and constraining the stability. This method, which is usually called the 
integrating factor method, allows an increase of the CFL number up to a factor
ten, but it cannot fix the dependence on $1/N^2$.

We do not have this difficulty because we use regularized dispersive terms.
The regularization effect can be seen from equation (\ref{eq:ubardiff}).
The same idea was exploited by \cite{Bona1981}, who used the modified KdV equation 
(\ref{eq:mKdV}).

Let us briefly explain how we treat the non-linear terms. Since the time integration scheme is explicit,
one can easily handle nonlinearities.
For example the term $\overline{(h+\eps\eta)\u}$ is computed as follows:
\begin{equation*}
	\overline{(h+\eps\eta)\u} = 
	\fft\left((h+\eps\Re\ifft(\overline\eta))\cdot\Re\ifft(\overline\u)\right).
\end{equation*}
The other nonlinear terms are computed in the same way.

\subsection{Validation of the numerical method}
One way to validate a numerical scheme is to compare the numerical results
with analytical solutions. Unfortunately, the authors did not
succeed in deriving analytical solutions to the $(1+1)$D dissipative Boussinesq equations
over a flat bottom. But for validation purposes, one can neglect the viscous term.
With this simplification several solitary wave solutions can be obtained.
We follow closely the work of \cite{Chen1998}.
In $(1+1)$D in the presence of a flat bottom, the Boussinesq system without dissipation becomes
\begin{equation}\label{eq:1deta}
  \eta_t + u_x + \eps(u\eta)_x - \frac{\m}{6}u_{xxx} = 0,
\end{equation}
\begin{equation}\label{eq:1du}
  u_t + \eta_x + \eps u u_x - \frac{\m}{2}u_{xxt} = 0.
\end{equation}
We look for solitary-wave solutions travelling to the left in the form
\begin{equation*}
  \eta (x,t) = \eta (\xi) = \eta (x_0 + x + ct), \qquad
  u(x,t) = B\eta(\xi),
\end{equation*}
where we introduced the new variable $\xi=x_0 + x + ct$ and $B$, $c$, $x_0$ are constants.
From the physical point of
view this change of variables is nothing else than Galilean transformation. In other
words we choose a new frame of reference which moves with the same celerity as 
the solitary wave. Since $c$ is constant (there is no acceleration), the observer
moving with the wave will see a steady picture.

In the following primes denote derivation with respect to $\xi$. Substituting
this special form into the governing equations (\ref{eq:1deta})-(\ref{eq:1du}) gives
\begin{equation*}
  c\eta' + u' + \eps (u\eta)' - \frac{\m}{6}u''' = 0,
\end{equation*}
\begin{equation*}
  cu' + \eta' + \eps uu' - c\frac{\m}{2}u''' = 0.
\end{equation*}
One can decrease the order of derivatives by integrating once:
\begin{equation*}
  c\eta + u + \eps u\eta - \frac{\m}{6}u'' = 0,
\end{equation*}
\begin{equation*}
  cu + \eta + \frac{\eps}{2}u^2 - c\frac{\m}{2} u'' = 0.
\end{equation*}
The solution is integrable on $\R$ and 
there are no integration constants, since a priori the solution 
behaviour at infinity is known: the solitary wave is exponentially 
small at large distances from the crest. Mathematically it can be
expressed as
\begin{equation*}
  \lim_{\xi\to\pm\infty} \eta(x,t) = \lim_{\xi\to\pm\infty} u(x,t) = 0.
\end{equation*}
Now we use the relation $u(\xi) = B\eta(\xi)$ to eliminate the variable $u$
from the system:
\begin{equation}\label{eq:comp1}
  (c+B)\eta - B\frac{\m}{6}\eta'' = -\eps B\eta^2,
\end{equation}
\begin{equation}\label{eq:comp2}
  (1+cB)\eta - cB\frac{\m}{2}\eta'' = -\frac{\eps}{2}B^2\eta^2.
\end{equation}
In order to have non-trivial solutions both equations must be compatible.
Compatibility conditions are obtained by comparing the coefficients 
of corresponding terms in equations (\ref{eq:comp1})-(\ref{eq:comp2}):
\begin{eqnarray*}
  \frac12 B^2 - \frac12 Bc &=& 1, \\
  \frac16 B^2 - Bc &=& 0.
\end{eqnarray*}
These relations can be thought as a system of linear equations with respect
to $B^2$ and $Bc$. The unique solution of those equations is
\begin{equation*}
  B^2 = \frac{12}{5}, \quad c = \frac{B}{6}. 
\end{equation*}
Choosing $B>0$ so that $c>0$ leads to 
$$ B=\frac{6}{\sqrt{15}}, \quad c = \frac{1}{\sqrt{15}}. $$
These constants determine the amplitude and the propagation speed of the solitary wave.
In order to find the shape of the wave, one differentiates once equation (\ref{eq:comp2}):
\begin{equation}\label{eq:givemesoliton}
  7\eta' - \m\eta''' = -12\eps \eta\eta'.
\end{equation}
The solution to this equation is well-known
(see for example \cite{Newell1977, Chen1998}):
\begin{lemma}
  Let $\alpha$, $\beta$ be real constants; the equation
  \begin{equation*}
    \alpha\eta'(\xi) - \beta\eta'''(\xi) = \eta(\xi)\eta'(\xi)
  \end{equation*}
  has a solitary-wave solution if $\alpha\beta > 0$. Moreover, the solitary-wave
  solution is 
  \begin{equation*}
    \eta(\xi) = 3\alpha\,\sech^2\left(\frac12\sqrt{\frac{\alpha}{\beta}}(\xi + \xi_0)\right)
  \end{equation*}
  where $\xi_0$ is an arbitrary constant.
\end{lemma}
Applying this lemma to equation (\ref{eq:givemesoliton}) yields the following solution:
\begin{equation}\label{eq:soliton}
  \eta(x,t) = -\frac{7}{4\eps}\,\sech^2\left(\frac{\sqrt{7}}{2\mu}(x + ct + x_0)\right),
\end{equation}
\begin{equation*}
  u(x,t) = -\frac{7\sqrt{15}}{10\eps}\,\sech^2\left(\frac{\sqrt{7}}{2\mu}(x + ct + x_0)\right).
\end{equation*}

Note that this exact solitary wave solution is not physical. Indeed the velocity is negative whereas
one expects it to be positive for a depression wave propagating to the left. In any case, the goal here
is to validate the numerical computations by comparing with an exact solution.
The methodology is simple. We choose a solitary wave
as initial condition and let it propagate during a certain time $T$ with the spectral
method. At the end of the computations one computes the  $L_\infty$ norm of the difference 
between the analytical solution (\ref{eq:soliton}) and the numerical one $\tilde{\eta}(x,T)$:
\begin{equation*}
  \epsilon_N := \max_{1\leq i \leq N}\abs{\eta(x_i,T) - \tilde{\eta}(x_i,T)},
\end{equation*}
where $\set{x_i}_{1\leq i \leq N}$ are the discretization points.

\begin{figure}
	\centering
		\includegraphics[width=0.8\textwidth]{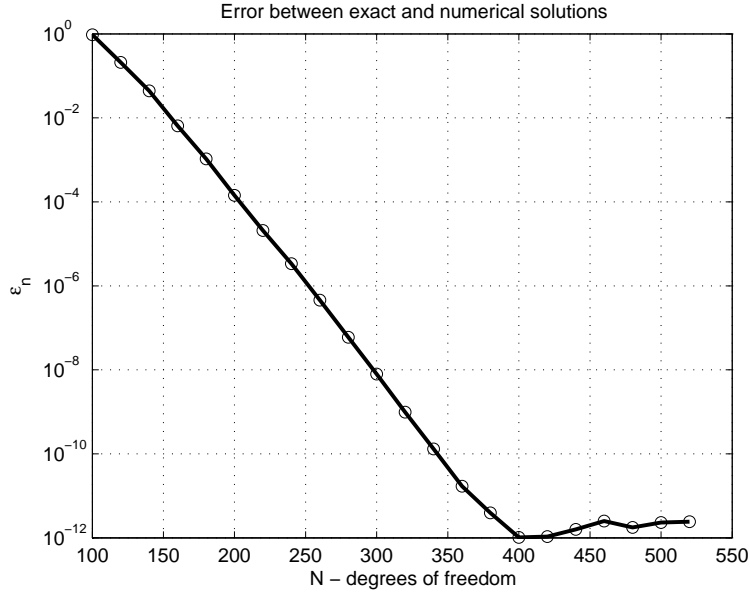}
	\caption{Error on the numerical computation of a solitary wave solution. Here $T=1$.}
	\label{fig:precision}
\end{figure}

Figure \ref{fig:precision} shows the graph of $\epsilon_N$ as a function of $N$. This result shows an excellent
performance of this spectral method with an exponential convergence rate. In general, the error
$\epsilon_N$ is bounded below by the maximum between the error due to the time integration algorithm 
and floating point arithmetic precision. 

The exponential convergence rate to the exact solution is one of the features of
spectral methods. It explains the success of these methods in several domains
such as direct numerical simulation (DNS) of turbulence. One of the main drawbacks
of spectral methods consists in the difficulties in handling complex geometries
and various types of boundary conditions.

\section{Numerical results}

In this section we perform comparisons between the two dissipation models
(\ref{eq:umodelI}) and (\ref{eq:umodelII}).
Even though the computations we show deal with a 1D wave propagating in the negative $x-$direction, 
they have been performed with the 2D version of the code. 
The bathymetry $z=-h(x,y)$ is chosen to be a regularized step function which is translated in the $y-$direction.
A typical function $h(x,y)$ is given by
\begin{equation}
	h(x) = \left\{
	\begin{array}{lc}
	  h_l,& x \leq x_0, \\
	  h_l + \frac12(h_r-h_l)\left(1+\sin\left(\frac{\pi}{\Delta x}(x-x_0-\frac12\Delta x)\right)\right), &
	  x_0 < x < x_0 + \Delta x, \\
	  h_r,& x \geq x_0 + \Delta x.
	\end{array}
	\right.
\end{equation}
%

This test case is interesting from a practical point of view since it clearly illustrates
the phenomena of long wave reflection by bottom topography. The parameters used in this computation 
are given in Table \ref{tab:numeric}. All values are given in nondimensional form.

\begin{table}
\begin{center}
\begin{tabular}{lccccccc}
  \hline
  {\it parameter} & $h_l$ & $h_r$ & $x_0$ & $\Delta x$ & $\eps$ & $\mu$ & $\nu_1$, $\nu_2$ \\
  \hline 
  {\it value} & 0.5 & 1.0 & $-0.5$ & 0.3 & 0.005 & 0.06 & 0.14 \\
  \hline
\end{tabular}
\caption[]{Typical values of the parameters used in the numerical computations}
\label{tab:numeric}
\end{center}
\end{table}

\subsection{Construction of the initial condition }
We propagate on the free surface a so-called approximate soliton. Its classical construction is as follows. 
We begin with the non-dissipative Boussinesq equations on a flat bottom:
\begin{equation*}
	\eta_t + \bigl((1+\eps\eta)u\bigr)_x - \frac{\m}{6} u_{xxx} = 0,
\end{equation*}
\begin{equation}
	u_t + \eta_x + \frac{\eps}{2}(u^2)_x - \frac{\m}{2} u_{xxt} = 0,
\end{equation}
and look for $u$ in the following form:
\begin{equation}\label{eq:relueta}
	u = -\eta + \eps P + \m Q + O(\eps^2 + \eps\m + \mm).
\end{equation}
It is precisely at this step that one makes an approximation.
One substitutes this asymptotic expansion into the governing equations and
retains only the terms of order $O(\eps+\m)$:
\begin{equation}\label{eq:p26_3}
	\eta_t - \eta_x + \eps P_x + \m Q_x - 2\eps\eta\eta_x + \frac{\m}{6}\eta_{xxx} = O(\eps^2 + \eps\m + \mm),
\end{equation}
\begin{equation*}
	-\eta_t + \eta_x + \eps P_t + \m Q_t + \eps\eta\eta_x + \frac{\m}{2}\eta_{xxt} = O(\eps^2 + \eps\m + \mm).
\end{equation*}
Add these two equations and set the coefficients of $\eps$ and $\m$ equal to 0:
\begin{eqnarray}
  \eps: & P_x + P_t - \eta\eta_x = 0,\label{eq:p27_5} \\
  \m:   & Q_x + Q_t + \frac16\eta_{xxx} + \frac12\eta_{xxt} = 0. \label{eq:p27_6}
\end{eqnarray}

Since the water depth is $h = 1 + \eps\eta = 1 + O(\eps)$, the approximate solitary wave should
travel to the left with a celerity $c = 1 + O(\eps)$ and depend on the variable $x+ct = x + t + O(\eps)$.
Consequently one has the following relations:
\begin{equation*}
	\pd{f}{t} = \pd{f}{x} + O(\eps+\m),
	\qquad f \in \{\eta, P, Q\}.
\end{equation*}
Replacing time derivatives by spatial ones in (\ref{eq:p27_5})-(\ref{eq:p27_6}) yields
\begin{equation*}
	P_x = \frac12\eta\eta_x, \qquad
	Q_x = -\frac13\eta_{xxx}.
\end{equation*}
By integration (using the fact that solitary 
waves tend to zero at infinity), one obtains
\begin{equation}
	P = \frac14\eta^2, \qquad Q = -\frac13\eta_{xx}
\end{equation}
and the relation (\ref{eq:relueta}) connecting $\eta$ and $u$ becomes
\begin{equation}\label{eq:uetarel}
	u = -\eta + \frac{\eps}{4}\eta^2 - \frac{\m}{3}\eta_{xx} + \ldots. 
\end{equation}
Substituting this expression for $u$ into (\ref{eq:p26_3}) yields
a classical KdV equation for $\eta$:
\begin{equation}
	\eta_t - \left(1+\frac32\eps\eta\right)\eta_x - \frac{\m}{6}\eta_{xxx} = 0,
\end{equation}
which admits solitary wave solutions of the form $\eta=\eta(x+ct)$:
\begin{equation*}
	\eta(x,t) = \frac{2(c-1)}{\eps}
	\sech^2\left(\frac{1}{2\mu}\sqrt{6(c-1)}(x+ct)\right),
\end{equation*}
where $c>1$. The velocity $u$ is obtained from (\ref{eq:uetarel}) by simple substitution.
This approximate soliton is used in the numerical computations.

\subsection{Comparison between the dissipative models}

\begin{figure}
	\centering
		\includegraphics[width=0.48\textwidth]{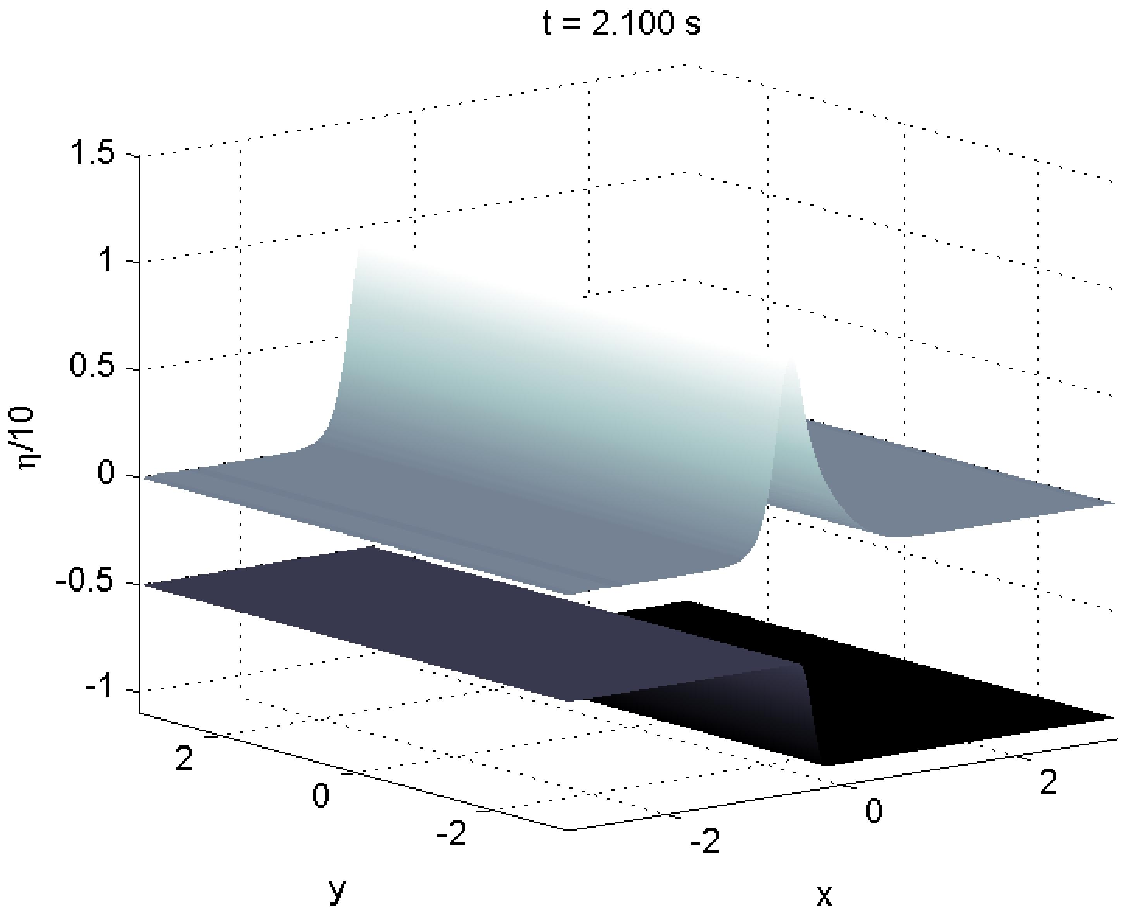}\includegraphics[width=0.48\textwidth]{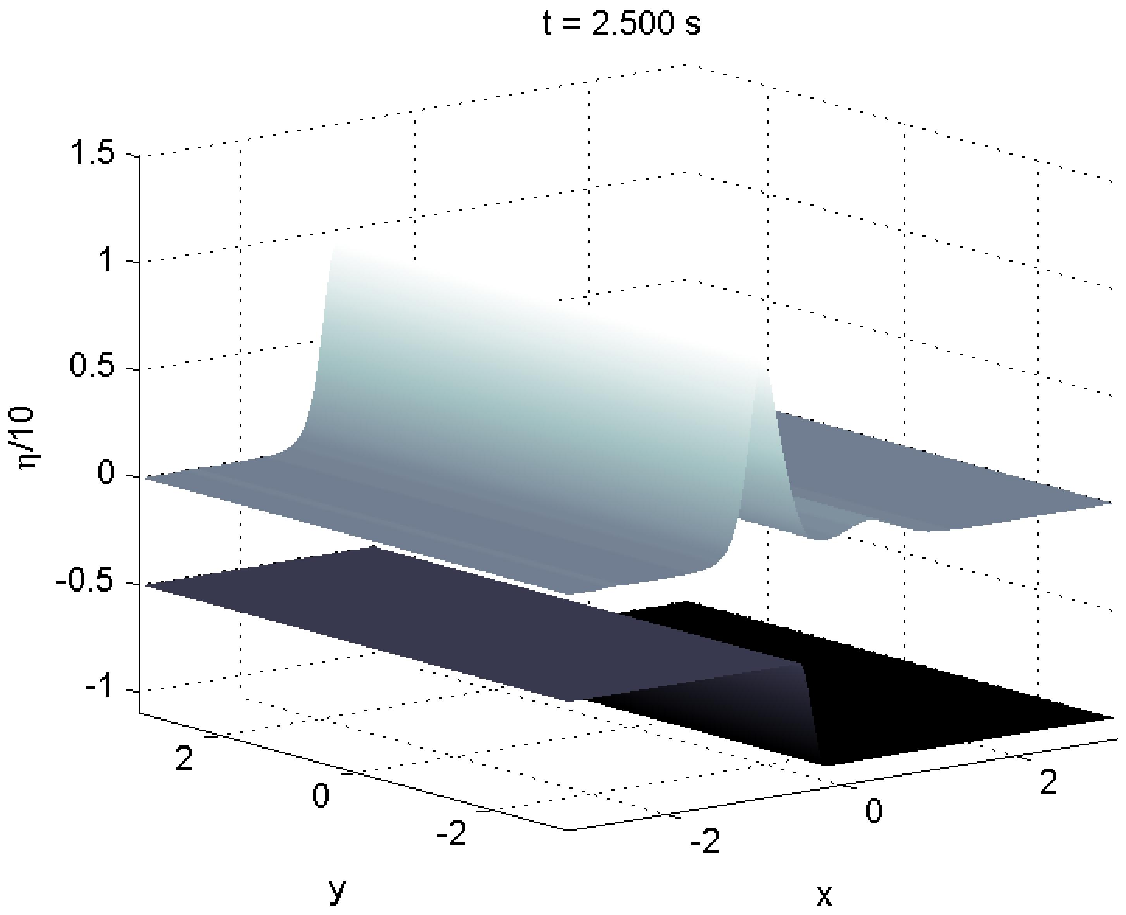}
	\caption{Interaction between a left-running solitary wave and a step at two different times. The plots represent $\eta/10$
while the true free-surface profiles are given by $z=\eps\eta$.}
	\label{fig:3dview}
\end{figure}

The snapshot of the function $\eta(x,y,t_0)$ (divided by 10 for clarity's sake) during and just after reflection by the step is given
on Figure \ref{fig:3dview}. Recall that the free surface is given by $z=\eps\eta$. 
Then we compare the two sets of equations 
(\ref{eq:etacommon}), (\ref{eq:umodelI}) and (\ref{eq:etacommon}), (\ref{eq:umodelII}). To do so we look at the 
section of the free surface at $y=0$ along the propagation direction.

\begin{figure}
	\centering
		\includegraphics[width=0.45\textwidth]{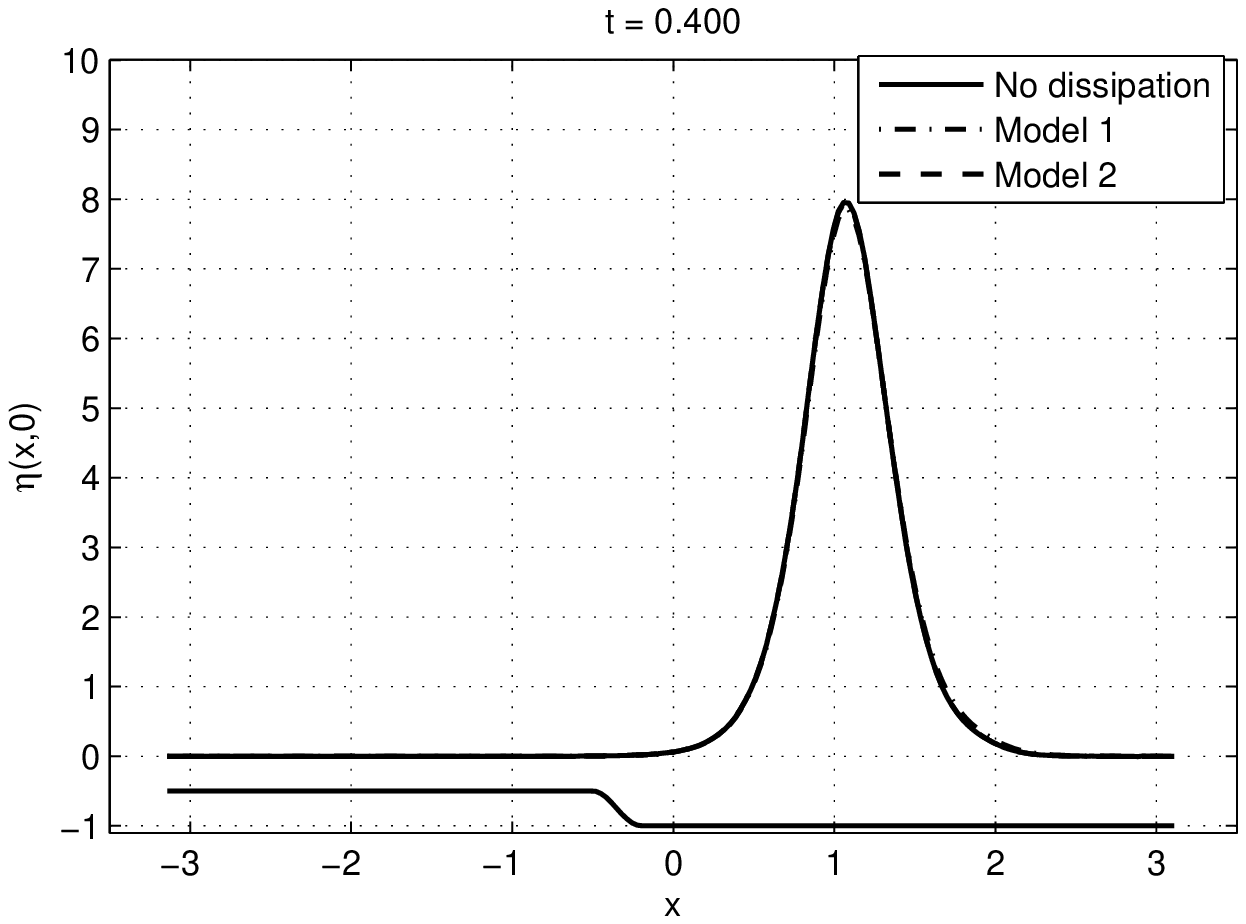}\includegraphics[width=0.45\textwidth]{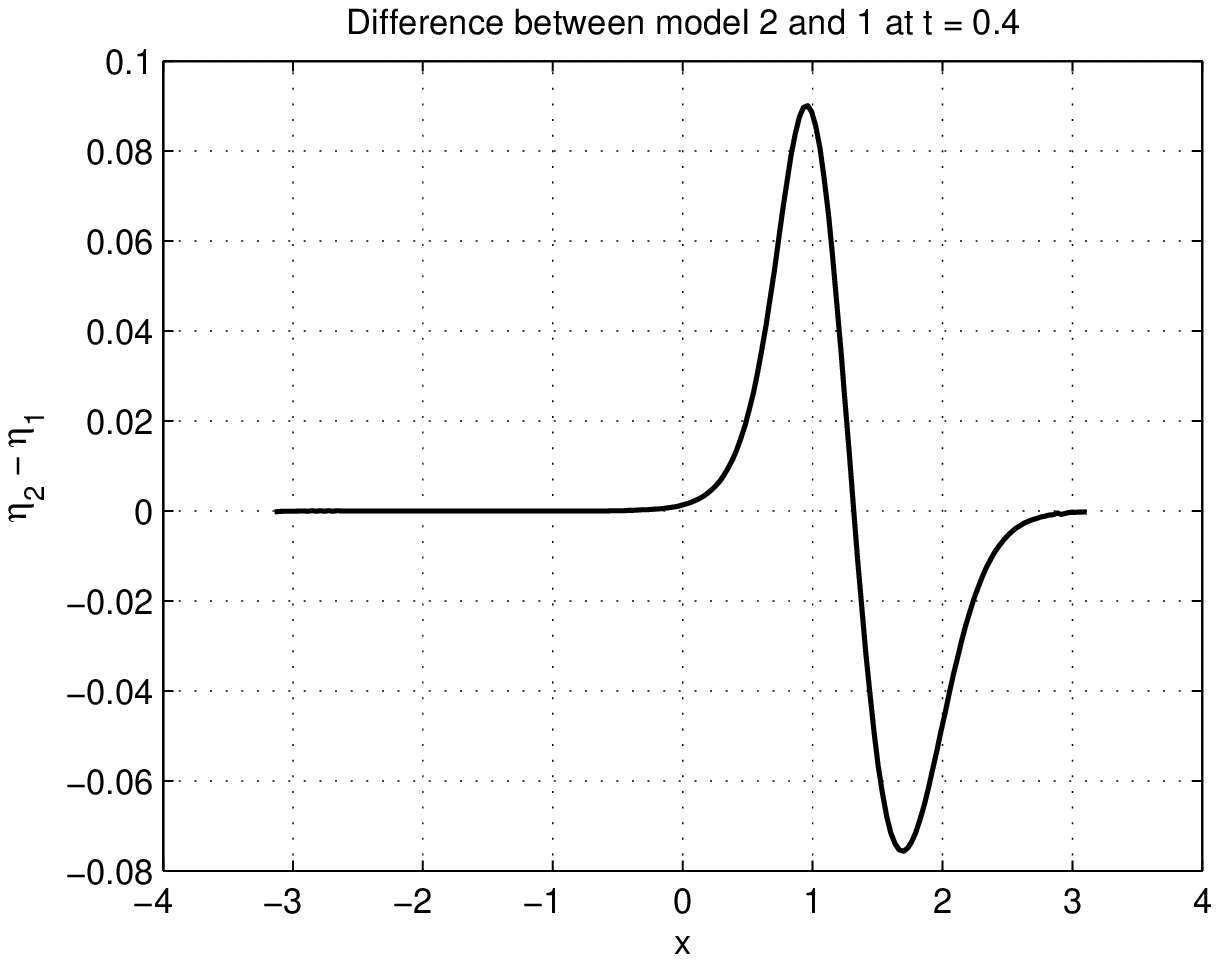}
	\caption{Free-surface snapshot before the interaction with the step: (left) the curves corresponding to the
three models are almost superimposed; (right) difference between model II and model I.}
	\label{fig:t0-39}
\end{figure}

Figure \ref{fig:t0-39} shows that even at the beginning of the computations the two
models give slightly different results. The amplitude of the pulse obtained with model I is smaller. It can be explained 
by the presence of the term $\nu_1S\u$ which is bigger in magnitude than $\eps\nu_2\nabla^2\u$. Within graphical accuracy, 
there is almost no difference between the conservative case and model II. 

\begin{figure}
	\centering
		\includegraphics[width=0.85\textwidth]{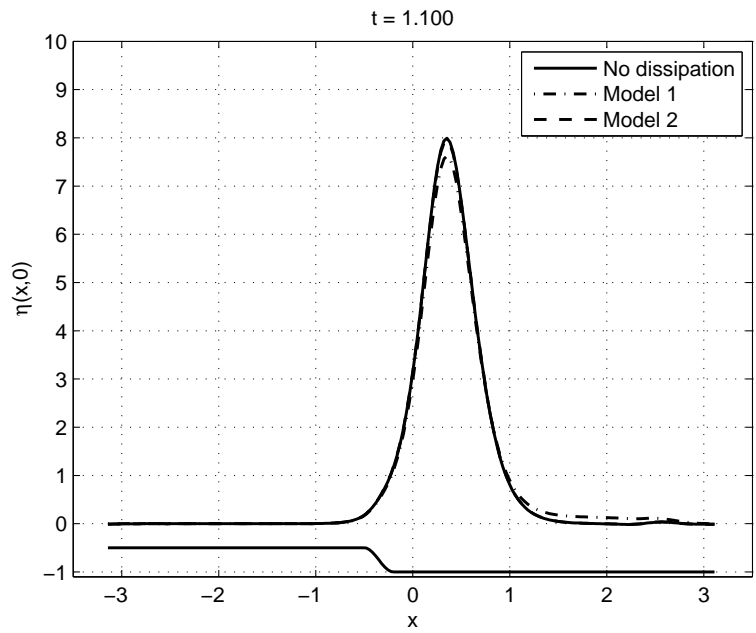}
	\caption{Free surface just before the interaction with the step}
	\label{fig:t1-08}
\end{figure}

In Figure \ref{fig:t1-08} one can see that differences between the two solitons 
continue to grow. In particular we see an important drawback of the dissipation
model I: just after the wave crest the free surface has some kind of residual
deformation which is clearly non-physical. Our numerical experiments 
show that the amplitude of this residue depends almost linearly on the parameter $\nu_1$.
We could hardly predict this effect directly from the equations without numerical
experiments.

\begin{figure}
	\centering
		\includegraphics[width=0.85\textwidth]{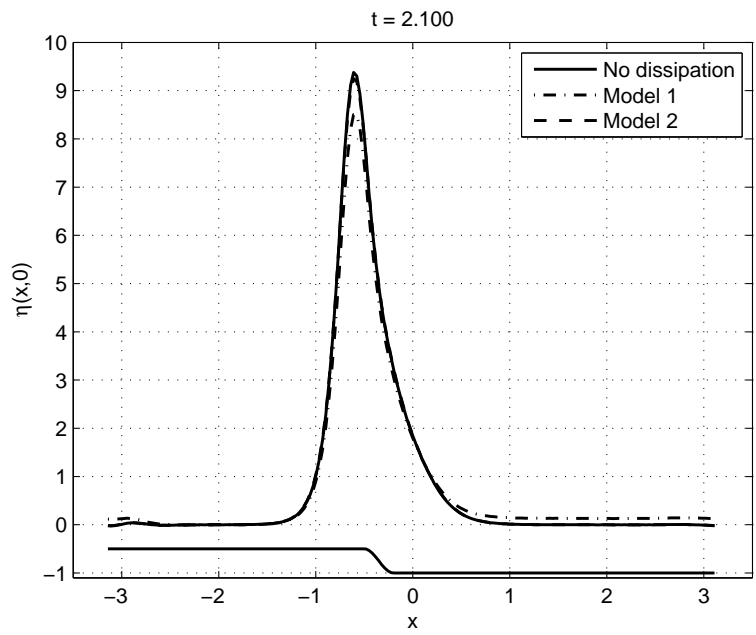}
	\caption{Beginning of the solitary wave deformation under the change in bathymetry}
	\label{fig:t2-10}
\end{figure}

We would like to point out several soliton transformations in Figure \ref{fig:t2-10}
due to the interaction with bathymetry. First of all, since the depth decreases, the wave amplitude grows.
Quantitatively speaking, the wave amplitude before the interaction is equal exactly to $8$
(without including dissipation) and over the step it becomes
roughly $9.4$. On the other hand the soliton becomes less symmetric which is also expected.
Because of periodic numerical boundary conditions we also observe the residue of the free-surface deformation coming
through the left boundary.

\begin{figure}
	\centering
		\includegraphics[width=0.85\textwidth]{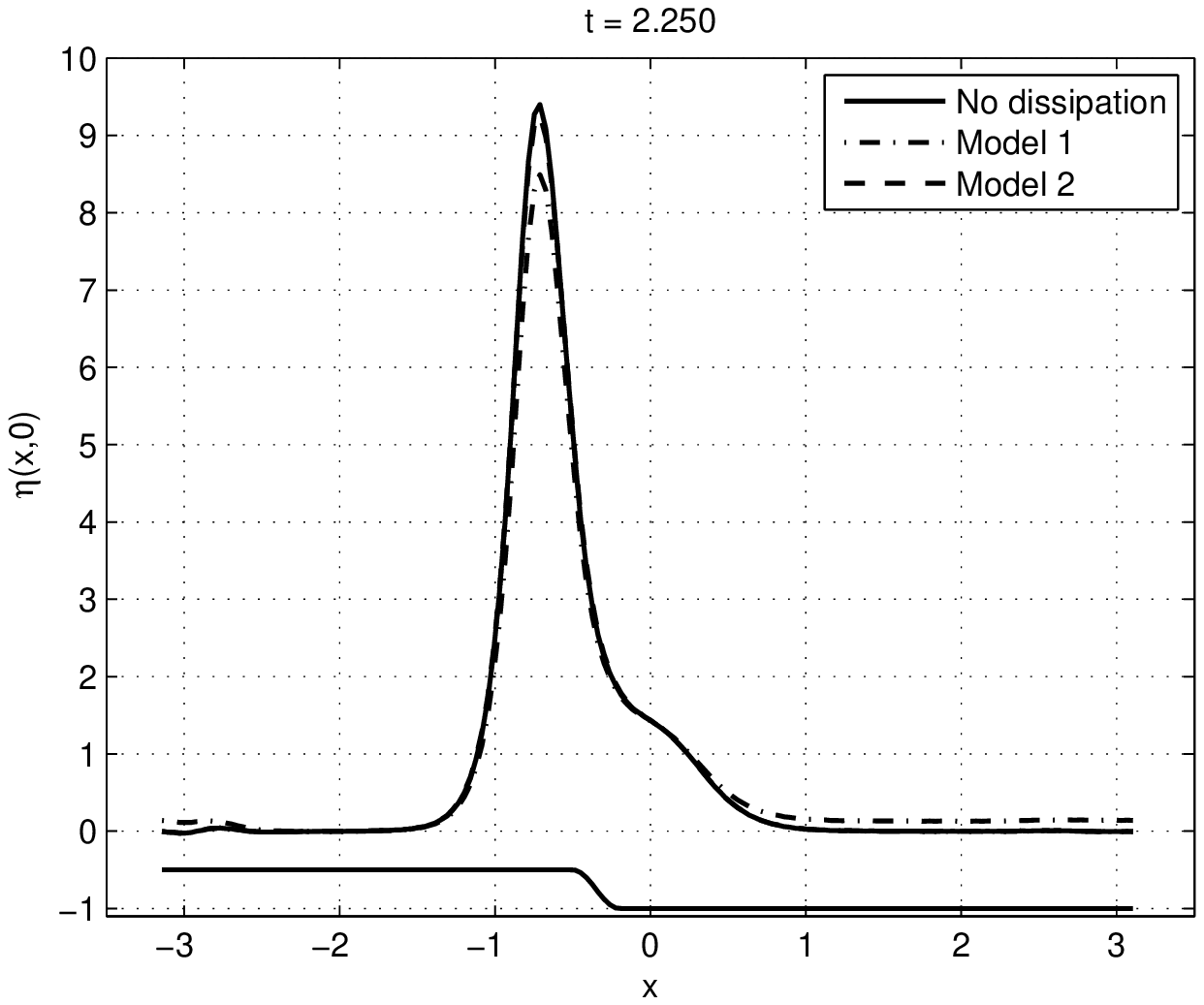}
	\caption{Initiation of the reflected wave separation}
	\label{fig:t2-24}
\end{figure}

\begin{figure}
	\centering
		\includegraphics[width=0.85\textwidth]{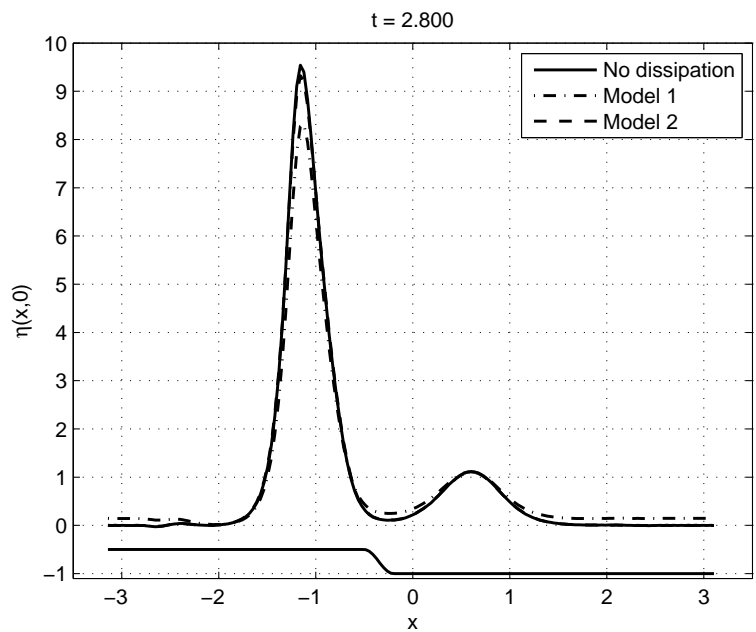}
	\caption{Separation of the reflected wave}
	\label{fig:t2-79}
\end{figure}

\begin{figure}
	\centering
		\includegraphics[width=0.85\textwidth]{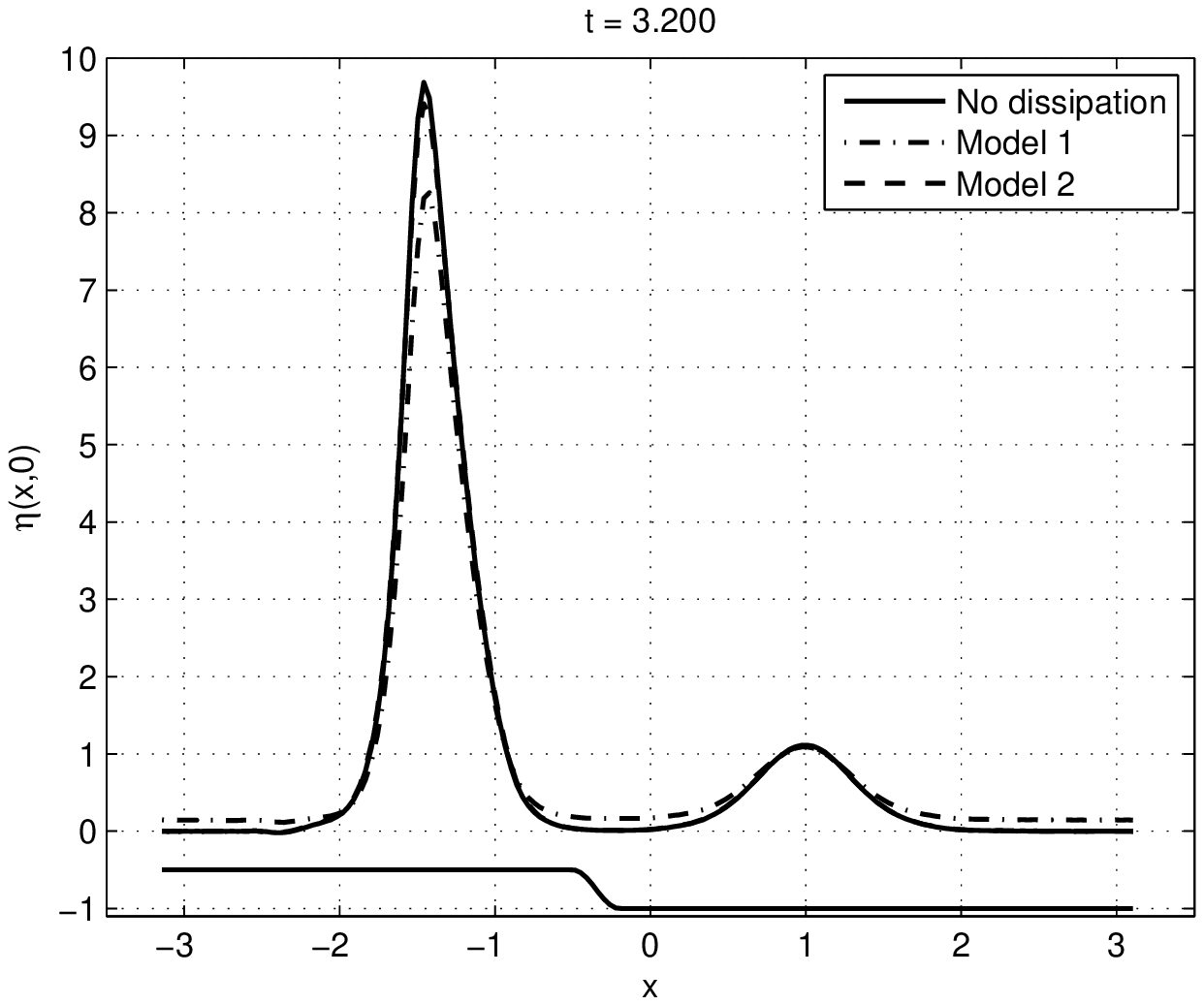}
	\caption{Two separate waves moving in opposite directions}
	\label{fig:t3-24}
\end{figure}

Figures \ref{fig:t2-24}, \ref{fig:t2-79} and \ref{fig:t3-24} show the process 
of wave reflection from the step at the bottom. The reflected wave clearly moves in the 
opposite direction. The fact that we see almost no difference between Model II and the conservative case
should not lead to the interpretation that dissipative effects are not important. One just has to wait long
enough to see these effects play a role. 

\section{Conclusions}
 
Comparisons have been made between two dissipation models. Model II, in which the decay is
proportional to the second derivative of the velocity, appears to be better.
At this stage we cannot show comparisons with laboratory experiments in order to
demonstrate the performance of model II. Nevertheless, there is an indirect evidence.
We refer one more time to the theoretical as well as experimental work of \cite{Bona1981}.
In order to model wave trains, they added to the Korteweg--de Vries equation an ad-hoc
dissipative term in the form of the Laplacian (but in 1D). This term coincides with the results of our derivation 
if we model dissipation in the equations according to the second model.
Their work shows excellent agreement between experiments and numerical solutions to
dissipative KdV equation. 
Moreover our dissipative Boussinesq equations are in the same relationship
with the classical Boussinesq equations \citep{Peregrine1967} as Euler and Navier-Stokes
equations. This is a second argument towards the physical pertinency of the results obtained with model II.


\bibliography{dissip}
\bibliographystyle{plainnat}
\end{document}